\documentclass[preprint,12pt]{elsarticle}

\usepackage{amssymb}
\usepackage{amsmath}
\usepackage[super]{nth}

\usepackage{float}

\usepackage{caption}
\usepackage{subcaption}
\biboptions{sort&compress}

\usepackage{xcolor}
\usepackage{tabularray}
\usepackage{natbib}
\usepackage{hyperref}
\hypersetup{colorlinks=true,
linkcolor=blue,
filecolor=magenta,      
urlcolor=cyan,
pdfauthor=author}

\usepackage{cleveref}
\usepackage{array}
\usepackage{booktabs}

\usepackage{eurosym}
\usepackage{siunitx}

\usepackage{csquotes}

\DeclareSIUnit{\EUR}{\text{\euro}}

\usepackage{xurl}

\usepackage{framed} 

\usepackage{multicol} 
\usepackage[super]{nth}

\usepackage{nomencl} 

\makenomenclature

\renewcommand*\nompreamble{\begin{multicols}{2}}

\renewcommand*\nompostamble{\end{multicols}}

\journal{Energy and AI}

\begin{document}

\begin{frontmatter}



\title{Real-world validation of safe reinforcement learning, model predictive control and decision tree-based home energy management systems}


\author[inst1]{Julian Ruddick\corref{cor1}}
\ead{julian.jacques.ruddick@vub.be}
\author[inst1,inst2,inst3]{Glenn Ceusters\corref{cor1}}
\ead{glenn.ceusters@be.abb.com}
\cortext[cor1]{Corresponding author}
\author[inst1]{Gilles Van Kriekinge}
\ead{gilles.van.kriekinge@vub.be}
\author[inst1]{Evgenii Genov}
\ead{evgenii.genov@vub.be}
\author[inst1]{Cedric De Cauwer}
\ead{cedric.de.cauwer@vub.be}
\author[inst1]{Thierry Coosemans}
\ead{thierry.coosemans@vub.be}
\author[inst1]{Maarten Messagie}
\ead{maarten.messagie@vub.be}

\affiliation[inst1]{organization={Electric Vehicle and Energy Research Group (EVERGI), Mobility, Logistics and Automotive Technology Research Centre (MOBI), Department of Electrical Engineering and Energy Technology, Vrije Universiteit Brussel},
            addressline={Pleinlaan 2}, 
            city={Brussels},
            postcode={1050},
            country={Belgium}}

\affiliation[inst2]{organization={ABB N.V.},
            addressline={Hoge Wei 27}, 
            city={Zaventem},
            postcode={1930}, 
            country={Belgium}}

\affiliation[inst3]{organization={AI lab, Vrije Universiteit Brussel},
            addressline={Pleinlaan 2}, 
            city={Brussels},
            postcode={1050},
            country={Belgium}}

\begin{abstract}
\sloppy Recent advancements in machine learning based energy management approaches, specifically reinforcement learning with a safety layer (\texttt{OptLayerPolicy}) and a metaheuristic algorithm generating a decision tree control policy (\texttt{TreeC}), have shown promise. However, their effectiveness has only been demonstrated in computer simulations. 
This paper presents the real-world validation of these methods, comparing them against model predictive control and simple rule-based control benchmarks.  
The experiments were conducted on the electrical installation of four reproductions of residential houses, each with its own battery, photovoltaic, and dynamic load system emulating a non-controllable electrical load and a controllable electric vehicle charger. 
The results show that the simple rules, \texttt{TreeC}, and model predictive control-based methods achieved similar costs, with a difference of only 0.6\%.
The reinforcement learning based method, still in its training phase, obtained a cost 25.5\% higher to the other methods.
Additional simulations show that the costs can be further reduced by using a more representative training dataset for \texttt{TreeC} and addressing errors in the model predictive control implementation caused by its reliance on accurate data from various sources.
The \texttt{OptLayerPolicy} safety layer allows safe online training of a reinforcement learning agent in the real world, given an accurate constraint function formulation. 
The proposed safety layer method remains error-prone; nonetheless, it has been found beneficial for all investigated methods. 
The \texttt{TreeC} method, which does require building a realistic simulation for training, exhibits the safest operational performance, exceeding the grid limit by only 27.1 Wh compared to 593.9 Wh for reinforcement learning.
\end{abstract}

\begin{graphicalabstract}
\includegraphics[width=\textwidth]{graphical_abstract.png}
\end{graphicalabstract}

\begin{highlights}
\item Constraint formulations can be decoupled from practical optimization applications;
\item A safety layer, \texttt{OptLayerPolicy}, allows to safely train RL agents online in the real-world;
\item Real-time safety layers can be beneficial for multiple (optimal) control methods;
\item Simple rules, machine learning and optimization based methods obtained equal costs;
\item Real test of a method generating an interpretable control strategy from data (\texttt{TreeC});
\end{highlights}


\end{frontmatter}


\begin{table*}[!t]   

    \begin{framed}
    
    \nomenclature{BESS}{battery energy storage system}
    \nomenclature{EV}{electric vehicle}
    \nomenclature{EMS}{energy management system}
    \nomenclature{MPC}{model predictive control}
    \nomenclature{PV}{photovoltaic}
    \nomenclature{RBC}{rule-based control}
    \nomenclature{RL}{reinforcement learning}
    \nomenclature{SOC}{state of charge}
    \printnomenclature
    \end{framed}
    
\end{table*}

\section{Introduction}
\label{sec:sample1}

Home energy management systems (EMS) are becoming increasingly relevant as we transition from fossil fuels towards further electrification of our energy systems. This transition is particularly notable with the adaption of affordable photovoltaic (PV) systems \cite{luderer_impact_2022}, which encourage renewable energy production at the household level. 
Alongside PV systems, controllable assets such as batteries, heat pumps and electric vehicles (EV) allow for optimized local consumption of this energy.
Electricity tariffs are also evolving \cite{hampton_customer_2022}, with dynamic pricing mechanisms on the household level being introduced in many countries to incentivise users to consume energy when it is abundant and cheap. 
These dynamic prices can vary within the same day, on the spot market, and across days, on the day-ahead market. 
It is not reasonable to expect a human to manually and consistently adjust their consumption to these volatile tariffs over long periods \cite{matisoff_review_2020}, especially given uncertainties like the significant variance of solar PV generation on partially cloudy days. 
Therefore, with the adoption of controllable electrical assets and dynamic prices, automated home EMSs become essential for optimizing economic or environmental objectives \cite{fabrizio_trade-off_2009} while ensuring user comfort.

Achieving and maintaining the near-to-optimal operation of any controllable system is a substantial undertaking. 
Adhering to all system constraints in real-time is always of primary concern (e.g., to avoid power outages in electrical systems or to avoid thermal discomfort in heating systems). 
Pursuing a specific objective is then only of secondary concern (e.g. minimizing energy costs or $CO_{2}$-equivalent emissions). 
In addition, different time dependencies could be required in the mathematical formulation, like with energy storage systems and controllable loads, which in principle would require an \textit{infinite} optimization horizon \cite{ceusters_model-predictive_2021}. 
All in the face of various uncertainties, such as fluctuating demands, weather conditions and prices, and so optimizing the \textit{expectation} in typically \textit{continuous} systems.

To tackle these complexities, model predictive control (MPC) has become a widely accepted and utilized optimal control technology in various industries \cite{forbes_model_2015}, including for building management \cite{drgona_all_2020}. MPC method is well-established and studied in terms of feasibility, stability, robustness and constraint handling \cite{gorges_relations_2017}. Nevertheless, it necessitates a detailed model \textit{a priori} (such as plant models, input/output disturbance models and measurement noise models), which is error-prone and usually lacks adaptability. Additionally, it might not be financially feasible to build, configure and maintain these models, especially in smaller systems \cite{sturzenegger_model_2016}. Recent advancements in machine learning approaches \cite{schwenzer_review_2021} have shown promising results in terms of modelling requirements, model convexity, adaptivity, constraint handling and system integration requirements. More specifically under consideration here, (1) using reinforcement learning (RL) in combination with a safety layer that can handle any constraint type and utilize an \textit{a priori} safe fallback policy when available (i.e., \texttt{OptLayerPolicy} \cite{ceusters_adaptive_2023}) and (2) using a metaheuristic algorithm to generate an interpretable control policy modelled as a decision tree (i.e., TreeC \cite{ruddick_treec_2023}). However, the effectiveness of these methods has only previously been shown in computer simulations. 

\subsection{Contribution and outline}

The original contributions outlined in this study are considered to be the first of their kind, to the best knowledge of the authors, and can be summarized as follows:

\begin{itemize}
    \item Real-world experimental validation of RL in combination with a safety layer that can handle any constraint type and utilize an \textit{a priori} safe fallback policy  (i.e., \texttt{OptLayerPolicy} \cite{ceusters_adaptive_2023}).
    \item Real-world experimental validation of a method using a metaheuristic algorithm to generate an interpretable control policy modelled as a decision tree (i.e., TreeC \cite{ruddick_treec_2023}).
    \item \sloppy Comparison between rule-based control (RBC), MPC, Safe (using \texttt{OptLayerPolicy}) RL and TreeC (on any case study, simulated or otherwise) using an evaluation procedure aimed to be as fair as possible.
\end{itemize}

In \cref{sec:literature}, we continue with a concise related work discussion, while \cref{sec:method} includes a detailed description of the utilized and proposed methodologies (i.e., hardware setup, safety layer, EMSs and evaluation procedure). 
\Cref{sec:results} then presents the results and a discussion of those results. 
Finally, \cref{sec:conclusion} presents our conclusions and formulates directions for future work.

\section{Related work}
\label{sec:literature}

Energy management systems (EMS) have been widely reviewed not only in terms of their mathematical modelling approach \cite{beaudin_home_2015}, but also considering their objectives, architecture, involved stakeholders and challenges \cite{rathor_energy_2020}, including for home EMSs more specifically \cite{zhou_smart_2016}. 
One observation, and also specifically highlighted by \citeauthor{rathor_energy_2020} \cite{rathor_energy_2020} and \citeauthor{allwyn_comprehensive_2023} \cite{allwyn_comprehensive_2023} in their reviews, is that most work is on simulated case studies and only a few show practical, real-world implemented results. 
Nevertheless, notable examples include \citeauthor{zhao_mpc-based_2015} \cite{zhao_mpc-based_2015} who implemented a non-linear MPC in the Hong Kong Zero Carbon Building that optimized the usage of an electric chiller and combined cooling and power unit, considering a stratified chilled water storage tank as thermal energy storage (and thus source of flexibility) and a PV system under a day-ahead pricing mechanism. 
\Citeauthor{viot_model_2018-1} \cite{viot_model_2018-1, viot_model_2018} implemented an MPC in the SYNERGI building of the University of Bordeaux - France, which optimized the Thermally Activated Building Systems and reported significant energy savings (40\%) with an improved or equivalent thermal comfort. \Citeauthor{arroyo_comparison_2022} \cite{arroyo_comparison_2022} then later compared white-, grey- and black-box modelling paradigms in an MPC-based home EMS for the Thermally Activated Building Systems of the Vliet building of the KU Leuven University in Heverlee, Belgium. \Citeauthor{elkazaz_energy_2020} \cite{elkazaz_energy_2020} implemented a two-layer structured EMS in a laboratory-based grid-connected microgrid, including a non-controllable load, a solar PV system, wind turbine production and a battery energy storage system (BESS). 
The mathematical program in the upper layer then determined the optimal reference values for the lower level MPC and reported a daily cost reduction of up to 30\%. 
\Citeauthor{zhang_model_2022} \cite{zhang_model_2022} then more recently presented the results of a year-long evaluation of an MPC for the energy management in a small commercial building, located in California, United States, equipped with solar PV and controllable space conditioning, commercial refrigeration, and a BESS while testing two types of demand flexibility applications: (1) electricity cost minimization under time-of-use tariffs and (2) responses to grid flexibility events. 
Their results showed 12\% annual electricity cost savings and 34\% peak demand reduction against a baseline, trained using supervised learning, while preserving thermal comfort and food safety. 
\Citeauthor{yang_experiment_2021} \cite{yang_experiment_2021} then also used supervised learning, yet to approximate an MPC and conducted an experimental study on two real testbeds (office floor and lecture theatre) in Singapore. 
The EMS controlled the air-conditioning systems and still managed to save 52\% of cooling energy consumption compared to 58\% of the original MPC but reduced the computational load by a factor of 100. 

When looking for real-world implemented machine learning EMS results, the options become even more scarce, as also mentioned by \citeauthor{fu_applications_2022} \cite{fu_applications_2022} in their review. Even though multiple real-world challenges are yet to be verified and thus remain open, as e.g. stipulated by \citeauthor{nweye_real-world_2022} \cite{nweye_real-world_2022}. 
This, together with recent advancements in imitation learning, makes it a more viable practical option \cite{dey_reinforcement_2023}. 
Nonetheless, notable implemented examples include \citeauthor{costanzo_experimental_2016} \cite{costanzo_experimental_2016} who demonstrated a model-assisted batch RL agent controlling heating, ventilation and air conditioning units, subjected to dynamic pricing, for the room climate control of a living lab and confirmed that within 10 to 20 days sensible control policies, that are further shaped by domain knowledge, could be obtained. 
\Citeauthor{kazmi_gigawatt-hour_2018} \cite{kazmi_gigawatt-hour_2018} then later demonstrated a model-based RL agent optimizing domestic hot water production, controlling reheating cycles of a domestic hot water storage vessel coupled to an air source heat pump in 19 real houses in the Netherlands. 
They reported a 20\% energy efficiency gain without any loss of user comfort (observed on a subset of 5 houses). 
Around the same time, \citeauthor{zhang_practical_2018} \cite{zhang_practical_2018} practically implemented and evaluated a deep RL agent for the optimal control of a radiant heating system in a real office building. 
The RL agent was trained offline on a physics-based model that was calibrated with real operational data before deploying online on the actual heating system (without taking additional exploratory actions). 
A heating demand savings of 16.6\% to 18.2\% was reported compared to an RBC strategy over a three-month-long evaluation period. Also considering distributed energy resources, \citeauthor{touzani_controlling_2021} \cite{touzani_controlling_2021} demonstrated a deep RL agent optimizing the heating, ventilation and air conditioning and BESS, in the presence of solar PV, of the FLEXLAB at Lawrence Berkeley National Laboratory in Berkeley, California, United States. 
The RL agent was also trained offline on a calibrated physics-based simulation model before the online deployment of the policy on the real system. 
They reported significant cost savings (up to 39.6\%) while maintaining similar thermal comfort levels over a three-week-long testing period. 
\Citeauthor{naug_deep_2022} \cite{naug_deep_2022} then presented a deep RL agent for a 5-zone simulated testbed, but also a real three-story building at the Vanderbilt University, United States - controlling heating setpoints in the air handling units for a full year. 
Here, again, the RL agent was trained offline and afterwards deployed online. 
If a performance degradation was detected, machine learning models of the building dynamics were updated and an offline relearning loop then also updated the RL agent (using these offline dynamic system models). 
As a final example, \citeauthor{gokhale_real-world_2023} \cite{gokhale_real-world_2023} conducted a 4-week experimental case study for the demand response of a cluster of 8 residential buildings using an RL-based energy coordinator together with a PI control-based power dispatcher - that also would override actions that could lead to thermal comfort violations, given they utilize the thermal mass of the buildings as cost-free source of flexibility.

While we have discussed a similar amount of real-world implemented related works, experimental machine learning energy management case studies are far less common. In almost all RL-based cases, the agents are safely trained offline, using detailed \textit{a priori} simulators (which may not be available or economically viable to build for each project - and can introduce significant modelling bias, leading to poor online performances). However, even without taking exploratory actions online, there are no guarantees that the deployed policy is safe across the entire state-action space. The few cases that do train the RL agents directly online have very simple overruling mechanisms in place that are not only highly invasive (e.g. compared to correcting the closest feasible actions) but also don't generalize well. Furthermore, while learnable decision tree-based EMS cases exist \cite{huo_decision_2021, dai_deciphering_2023} - they have, to the best of our knowledge, never been demonstrated in the real world. Therefore, we filled in this research gap by (1) using an online RL-based EMS in combination a safety layer that can handle any constraint type and utilize an \textit{a priori} safe fallback policy when available (i.e., \texttt{OptLayerPolicy} \cite{ceusters_adaptive_2023}) and (2) using a learned interpretable decision tree based EMS (i.e., TreeC \cite{ruddick_treec_2023}), in the real-world. 

\section{Method}
\label{sec:method}

\subsection{Hardware setup}
\label{sec:hardware}

The experimental setup consists of four real reproductions of a house's electrical circuit. Each house has a BESS, a PV installation, an emulated electric load consumption and emulated EV charging sessions. The houses are connected to the grid and have official digital meters from the local Belgian distribution system operator that measures energy imported from and exported to the grid. The BESS and PV installations are different for each house, whereas the emulated electric load consumption and charging sessions are identical for all houses. This setup replicates an electrical configuration typical of a modern home with any non-electrical heating system.

The electrical loads used for the consumption and charging sessions come from a dynamic load system composed of seven 2000W Perel\textregistered PHP2000 heaters controlled through Showtec Single DP-1 dimmers.
The electrical loads are calibrated so that the setpoints match the real measured active power consumption more closely (we expect this difference to come from measurement\footnote{No voltage and therefore power factor, measurement \textit{per} circuit was operational} inaccuracies and manufacturing tolerances of the loads). 

Measurement devices for power, voltage and current are present for the PVs, BESSs and grid connections.
The power of the electrical loads and other power losses in the setup is calculated via the PV, BESS and grid measurements.

\begin{figure}[H]
    \centering
    \includegraphics[scale=0.6]{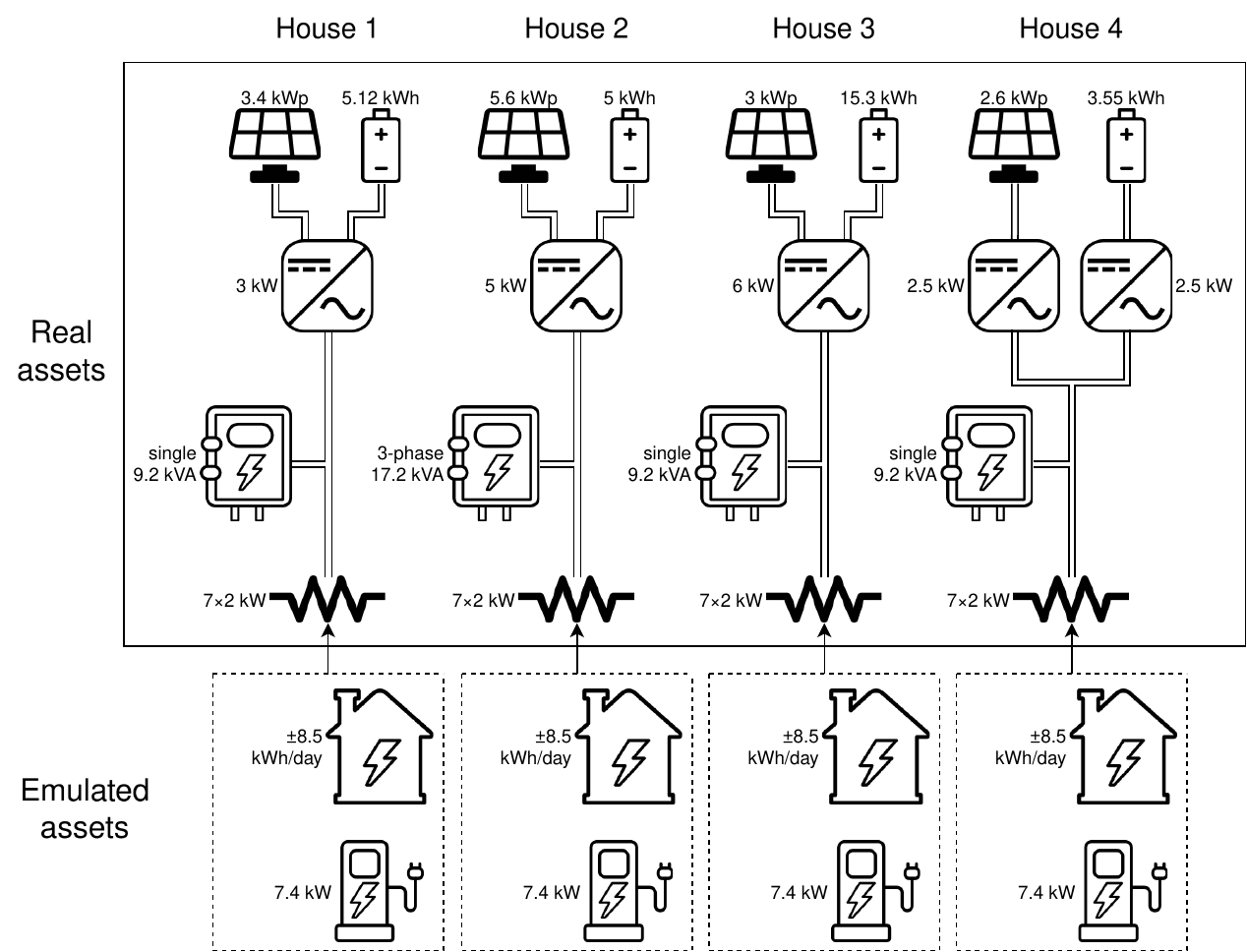}
    \caption{Schematic of the experimental setup per house with from top to bottom: 
the PV installation with its peak power, 
the BESS with its capacity, 
the inverter with its maximum power, 
the grid connection with its phase specification and maximum power, 
the dynamic loads with their maximum power, 
the emulated house consumption with its average daily consumption and 
the EV charger with its maximum charge power.}
    \label{fig:setup_diagram}
\end{figure}

\begin{figure}
    \centering
    \includegraphics[width=0.9\textwidth]{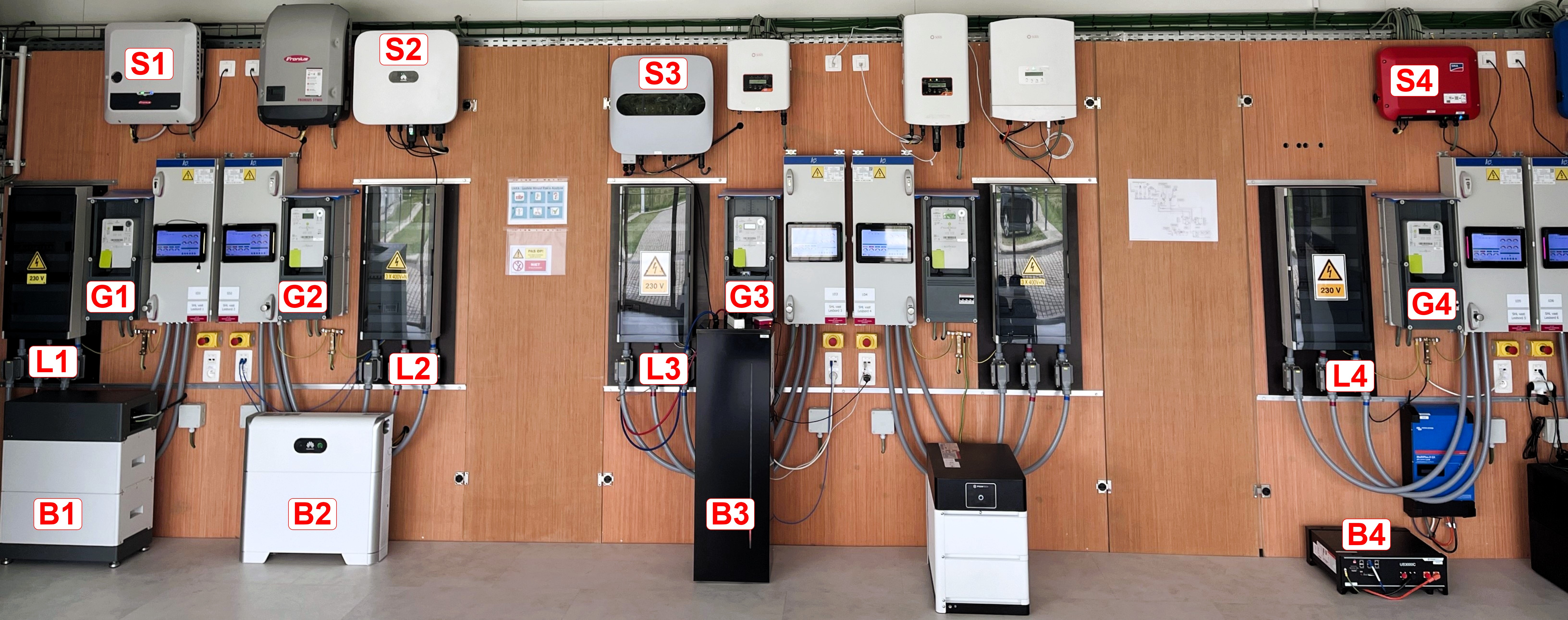}
    \caption{Photograph of the experimental setup: with \texttt{G} being the grid connection equipped with the official digital meter of the local distribution system operator, \texttt{B} the BESS, \texttt{S} the solar PV installation inverters (solar panels equipped on the roof) and \texttt{L} the connectors to the dynamic loads (dimmers in separate cabinet and electrical loads stationed outdoors). The number represents the house numbers.
    }
    \label{fig:svl_picture}
\end{figure}

A schematic and a picture of the setup are displayed in \cref{fig:setup_diagram,fig:svl_picture}. Additional hardware specifications for the BESSs, inverters, PVs and electrical circuits of each house are described in \cref{tab:batt_spec,tab:inverter_spec,tab:pv_spec}.

\subsection{Experimental setup}

\subsubsection{Consumption profile}
\label{sec:consumption}
We utilize the real-time (10 seconds) residential electricity \enquote{household} consumption data from \citeauthor{schlemminger_dataset_2022} \cite{schlemminger_dataset_2022}, focusing specifically on the 2019 dataset, which predates the COVID-19 pandemic. Our experiments employ data from single-family house (SFH) 19, which does not include solar PV, BESS, or EV charging. Additionally, this household dataset excludes heat pump loads. SFH 19 is selected due to its high data availability and quality, along with its typical consumption metrics, averaging 3425 kWh per year.

\begin{figure}[H]
    \centering
    \begin{subfigure}{0.45\textwidth}
        \centering
        \includegraphics[width=\linewidth]{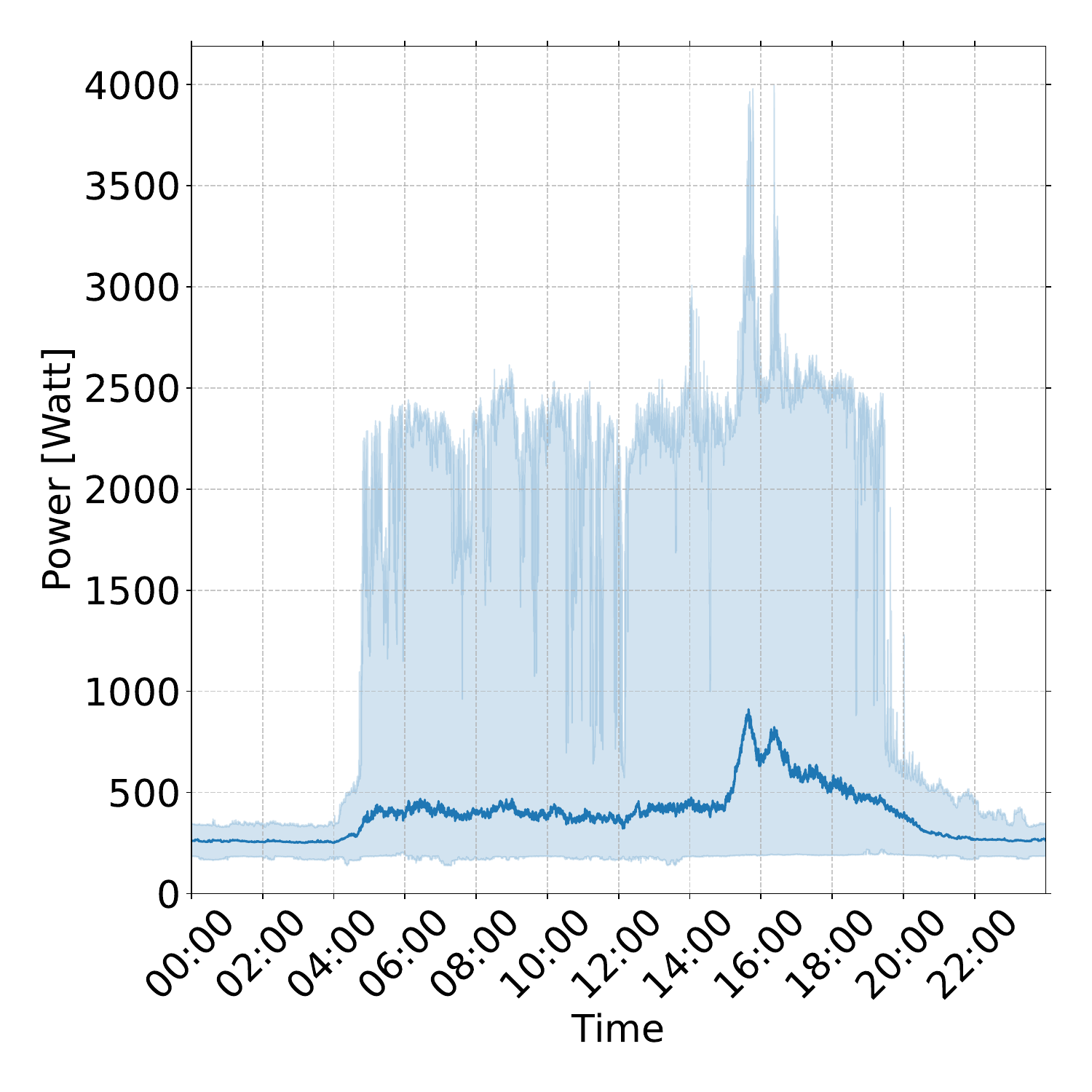}
        \caption{Time series with 2.5th and 97.5th percentile}
        \label{subfig:load lineplot}
    \end{subfigure}
    \hfill
    \begin{subfigure}{0.45\textwidth}
        \centering
        \includegraphics[width=\linewidth]{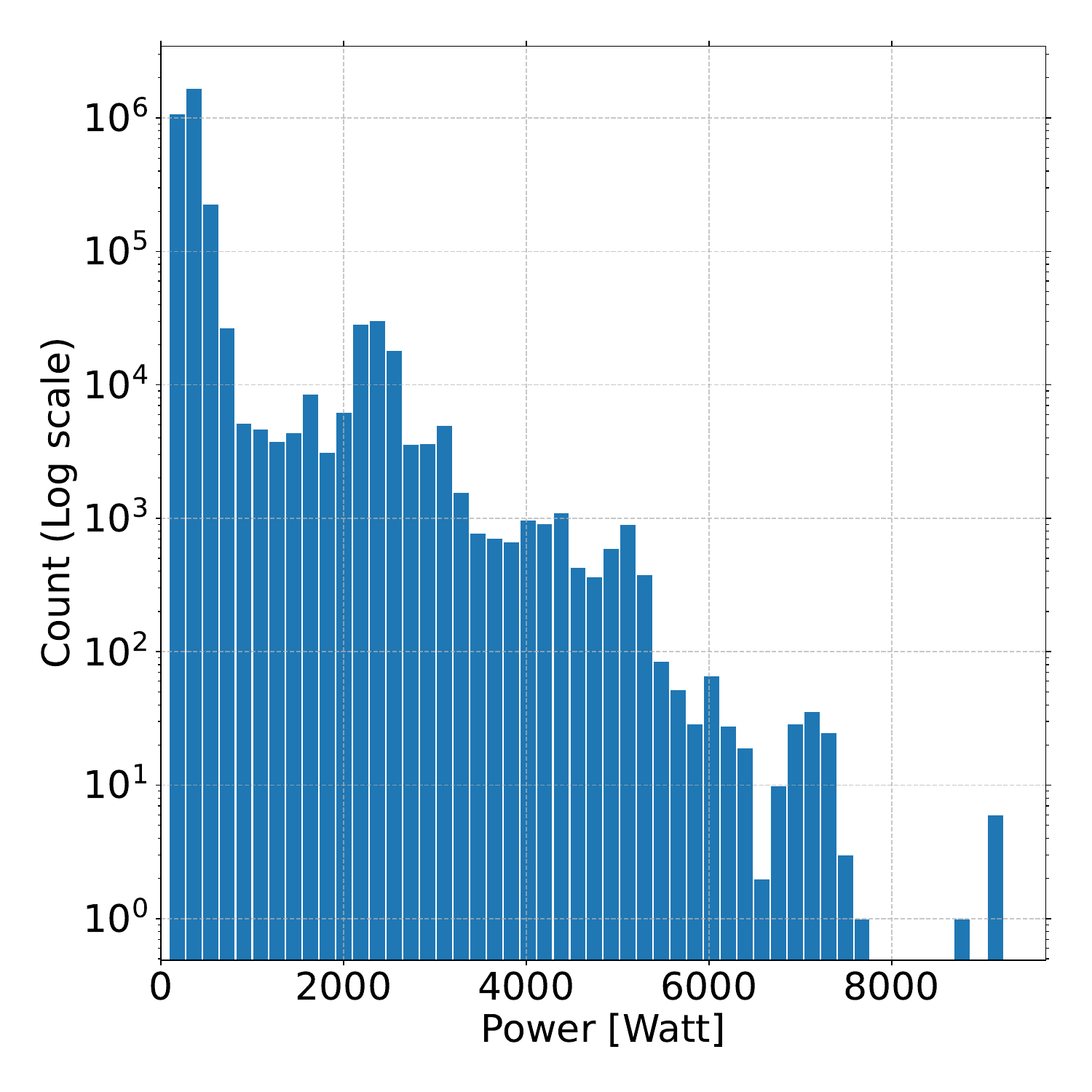}
        \caption{Histogram}
        \label{subfig:load histogram}
    \end{subfigure}
    \caption{Household consumption profile: 10 seconds SFH 19 data \cite{schlemminger_dataset_2022} without PV, BESS, EV charging or heat pump having 3425.75 kWh/year, Std: 445.85 W, max: 9229 W.}
    \label{fig:consumption profile}
\end{figure}

We shifted the dataset so that weekdays align (that a Monday in runtime is also a Monday in the dataset) and used this for all our hardware-in-the-loop simulated dynamic loads so that each house has the same real-time non-controllable load.

\subsubsection{Electric vehicle and driver charging behaviour}
\label{sec:ev_sessions}
The residential electric mobility demand consists of two parts: a) a first part specific to the EV user charging behaviour (i.e. arrival time, departure time and energy need), and b) a second part specific to the EV charging power profile.

Firstly, the EV user charging behaviour is simulated directly using an existing open-source dataset available in \cite{sorensen_residential_2021}. This dataset comprises data from 97 EV users, encompassing 6878 charging sessions at residential locations spanning from December 2018 to January 2020. 
Among these users, \enquote{Bl2-2} has been selected for the experiment due to its alignment with the average EV user in terms of arrival time, parking duration, and energy requirements. 
However, it distinguishes itself by exhibiting a notably high frequency of charging sessions, rendering it particularly interesting for the experiment as it allows for more interactions with the EMS developed in this paper. From a statistical standpoint, this EV user has undergone 315 charging sessions, with an aggregate energy consumption of 3274.74 kWh across 386 days.

Secondly, the EV charging power profile is built based on certain variables and constraints. The EV charging power is constrained by (\ref{eq_ev_power_charge}).
\begin{equation}
    0 \leq P^{ev}_{t} \leq P^{ev}_{max}
\label{eq_ev_power_charge} 
\end{equation}
where $P^{ev}_{max}$ is the maximum charging power in $[kW]$ set to 7.4 kW, following the IEC 61851 charging standards \cite{61851}. 
Additionally, the maximum charging power follows the Constant-Current/Constant-Voltage (CC-CV) uncoordinated charging method defined in \citeauthor{vagropoulos_optimal_2013} \cite{vagropoulos_optimal_2013}. Such a method stipulates that the power is a function of the state of charge (SOC) as a piece-wise linear function summarized in (\ref{ch2_sc3_eq_p_final_soc}).
\begin{equation}
P^{ev}_t = \left\{
\begin{aligned}
&P^{ev}_{max}   &\textrm{if} \quad SOC^{ev}_{t} \leq SOC^{ev}_{cc,cv}  \\
&P^{ev}_{max}  - (P^{ev}_{max}  - P^{ev}_{min}) * \frac{SOC^{ev}_{t} - SOC^{ev}_{cc,cv}}{1 - SOC^{ev}_{cc,cv}}  &\textrm{if} \quad SOC^{ev}_{t} > SOC^{ev}_{cc,cv} 
\end{aligned}
\right.
\label{ch2_sc3_eq_p_final_soc} 
\end{equation}
where $SOC^{ev}_t$ is the SOC at time t in $[\%]$, 
$SOC^{ev}_{cc,cv}$ is the transition SOC between the Constant-Current phase and the Constant-Voltage phase in $[\%]$ and $P^{ev}_{min}$ is the maximum charging power at 100\% SOC, set to 1 kW. Finally, both charging power and SOC variables are linked using the equality constraint (\ref{eq_ev_soc_balance}).
\begin{equation}
SOC^{ev}_{t+1} = SOC^{ev}_{t} + \eta^{charge} \times \frac{P^{ev}_{t} \times \Delta t}{E^{ev}}  \label{eq_ev_soc_balance} 
\end{equation}
where $SOC^{ev}_{t+1}$ is the expected SOC at time step $t+1$, $\Delta t$ is the time step length in $[h]$, $E^{ev}$ is the capacity of the EV in $[kWh]$ set to 60 kWh and $\eta^{charge}$ is the charging efficiency in $[\%]$ set to 95\% which is typical for lithium-ion batteries \cite{farhad_introducing_2019}.

\subsubsection{Tariff}
The real-time pricing used in this study is based on a one-year dynamic contract of a major Belgian energy supplier, as if it were taken in April 2024\footnote{Full contract available on the paper's GitHub repository at \url{https://github.com/EVERGi/real_validation_saferl_treec_paper/}}.
The total cost is composed of four costs with different pricing mechanisms: the day-ahead cost, the offtake extras cost, the peak cost and the fixed yearly cost.

The day-ahead cost depends on the prices of \cref{eq:injection_price,eq:offtake_price}.
\begin{equation}
    Vo_t= 
\begin{cases}
    (Vd_t+0.011\frac{\SI{}{\EUR}}{kWh})*1.06& \text{if } (Vd_t+0.011\frac{\SI{}{\EUR}}{kWh})\geq 0\\
    (Vd_t+0.011\frac{\SI{}{\EUR}}{kWh}),              & \text{otherwise}
\end{cases}
\label{eq:injection_price}
\end{equation}

\begin{equation}
    Vi_t=Vd_t-0.009\frac{\SI{}{\EUR}}{kWh}
\label{eq:offtake_price}
\end{equation}
Where $Vo_t$ and $Vi_t$ are the offtake and injection prices at time step $t$.
$Vd_t$ is the Belgian day-ahead hourly price.
A value-added tax of 6\% is added to the offtake price when it is positive.
Extra costs that occur for energy taken from the grid for transport, distribution and taxes are calculated together and labelled as offtake extras cost.
The offtake price for all these additional offtake costs $Vf$ is 0.114 \euro/kWh. 

The monthly peak offtake cost $Cp_m$ is calculated by \cref{eq:peak_cost}.
\begin{equation}
Cp_m = \frac{|B_m|}{N_m}Vp*\max_{{t\in B_m}} (\frac{Eo_t}{0.25h}, 2.5 kW)
\label{eq:peak_cost}
\end{equation}
Where $Vp$ is the peak price of 3.5 \euro/kW. 
This cost adds a penalty to the highest offtake power averaged over a 15-minute time step that occurred for each month. 
In the context of this study, the monthly peak offtake cost is proportional to the billed period of each month to also include this cost for unfinished months. 
$B_m$ is the set containing all billed time steps in month $m$ and $N_m$, which is the total number of time steps in the month.
$Eo_t$ is the energy taken out of the grid at time step $t$.
The offtake energy is divided by the length of a time step to obtain the power in kW.
A minimum monthly peak power of 2.5 kW is enforced in the cost, even if the measured peak power is lower.

The fixed yearly cost is calculated using the yearly price $Vy$ of \euro 115.84 and is proportional to the billing period.
The total electricity cost is calculated by \cref{eq:tot_cost}.

\begin{equation}
C_{tot}=\underbrace{\sum_{t\in T}{(Eo_tVo_t-Ei_tVi_t)}}_{\text{day-ahead cost}}+\underbrace{\sum_{t\in T}Eo_tVf}_{\text{offtake extras cost}}+\underbrace{\sum_{m \in M}{Cp_m}}_{\text{peak cost}}+\underbrace{\frac{|T|}{N_y}Vy}_{\text{yearly cost}}
\label{eq:tot_cost}
\end{equation}
Where $T$ is the set of all billed time steps, $Ei_t$ is the energy injected to the grid at time step $t$, $M$ is the set of billed months, and $N_y$ is the total number of time steps in the year.

\subsubsection{House switching}
As described in \cref{sec:hardware}, the four houses have different BESS and PV installations. 
Assigning a different EMS to each house and comparing the performance by running them simultaneously would not yield comparable results.
One house could obtain better performances because it has a better BESS or PV installation.
To avoid this bias, the EMSs are switched every 24 hours to a different house.

Since there are 4 houses and 4 EMSs to test, there are a total of 24 possible house and EMS combinations. 
Each combination is tested twice, resulting in 48 days of testing in total. 
The 48-day schedule is obtained by randomly shuffling the 24 combinations until two distinct 24-day schedules are generated. This ensures no consecutive days have the same reinforcement learning or MPC EMS assigned to the same house. 
This rule is applied for practical reasons, as the RL and MPC EMSs behaviours depend on how they operated a house previously. 
Failed test days can occur, and not having an MPC or RL EMS on the same house for consecutive days provides more time to identify a failure while preventing its propagation to the following day.

The switching is conducted daily at 15:00. By this time, the BESSs of all houses are charged to 100\% to ensure consistent starting conditions post-switch. 
The EV schedule is adjusted to prevent charging sessions from overlapping with the 15:00 switching time. 
This is achieved by shortening the duration of overlapping sessions, either by starting or finishing them at 15:00, with the option of the longest duration being selected. 
The start and end SOC of the EV batteries are maintained according to the original schedule.

These switching rules have been selected because 15:00 is a time when there are fewer charging sessions compared to the morning and evening hours. 
At 15:00, the BESS also had the opportunity to charge using the excess energy generated by the PV installation in the preceding hours.
Additionally, at this time, the day-ahead prices for the next day are available, eliminating the need to forecast them for the MPC EMS.

\subsubsection{Enforced charging behaviour}
\label{sec:enforced_charging}

An enforced charging behaviour is enforced in order to reach 100\% SOC at the switching time for the BESSs and to reach the imposed final SOC of an EV battery at the end of a charging session.
The batteries are charged at maximum power to reach their SOC goal at the latest possible time minus a buffer time $Btime$.
This buffer time is used to avoid not reaching the SOC goal in time due to non-perfect battery models, safety layer activation (which could reduce the battery setpoints, see \cref{sec:safety}) or other unexpected event.

The buffer time value is higher when the difference between the SOC goal and the current SOC of the battery is higher because more events could prevent the battery from reaching its SOC goal in time. This relation is described in \cref{eq:time_buffer}.

\begin{equation}
    B_{time} = (SOC_g-SOC)*(B_{max}-B_{min})+B_{min}
    \label{eq:time_buffer}
\end{equation}
Where $B_{time}$ is the time buffer, $SOC_g$ is the SOC goal of the battery, $SOC$ is the current SOC of the battery, $B_{max}$ is the time buffer enforced if 100\% of the battery still needs to be charged and $B_{min}$ is the minimum enforced time buffer.
Every 5 seconds, the battery controller calculates with equations \cref{eq_ev_soc_balance,eq:batt_model} whether the battery can reach the SOC goal at the intermediate time, defined as the target time minus the time buffer. 
If not, the battery is charged at maximum power for 1 minute.
For both the BESSs and the EV batteries, $B_{min}$ is set to 3 minutes. 
For the BESSs $B_{max}$ is set to 1 hour and for the EV batteries $B_{max}$ is set to 4 hours. 
The $B_{min}$ and $B_{max}$ values were obtained by testing in simulation what values would consistently meet the SOC goal in time throughout the 3 months of training data (see \cref{sec:trainingdata}).

\subsection{Safety layer}
\label{sec:safety}
Given that we are proposing to demonstrate native unsafe energy management methods, we implement the \texttt{OptLayerPolicy} safety layer from \citeauthor{ceusters_adaptive_2023} \cite{ceusters_adaptive_2023} - as it is considered one of the state-of-the-art methods. 
This safety layer results in the minimally invasive correction of predicted (control) actions towards their closest feasible, safe set - without affecting optimality. 
One can then also compute the distance from the feasible solution space and fallback on an \textit{a priori} safe policy when available. 
This is under the notion that far-from-feasible actions are likely to be far-from-optimal, while close-to-feasible can mean close-to-optimal. 
This threshold is a hyperparameter of \texttt{OptLayerPolicy} specifically, and its effectiveness was shown in a simulated case study by \citeauthor{ceusters_adaptive_2023} \cite{ceusters_adaptive_2023}.

While the safety layer mainly serves the RL-based EMS, it provides general constraint satisfaction guarantees, including for the simple RBC EMS benchmark (which is not necessarily always safe). 
We chose the same sampling rate for the safety layer as for the hardware-in-the-loop sampling rate, which was 5 seconds. 
By doing so, we effectively utilize it additionally as a real-time optimization step in a cascading (a.k.a. hierarchical) system architecture \cite{scattolini_architectures_2009} - including for the MPC-based EMS. 
Hence, it allows us to use slower sampling rates inside the EMSs (i.e., 15 minutes), where the optimized setpoints are then used by the safety layer as a reference signal (only deviating from it if it is absolutely necessary to subject to the constraints - considering the latest measurements, e.g. when the solar PV in-feed suddenly drops).

In our experimental case study, the \texttt{OptLayerPolicy} safety layer is expressed as: 

\begin{subequations}
\begin{align}
    \label{equation8a}
    a^{safe, bess}_{t}, a^{safe, ev}_{t}  &= \arg\min_{a^{bess}_{t}, a^{ev}_{t}} \frac{1}{2} \lVert (a^{bess}_{t} - \tilde{a}^{bess}_{t}) + (a^{ev}_{t} - \tilde{a}^{ev}_{t}) \rVert^2 \\[1em]
    \label{equation8b}
    \text{s.t.} \quad & -P^{grid}_{max} \leq P^{grid}_{t} \leq P^{grid}_{max}  &&\forall t\\
    \label{equation8c}
    & 0 \leq P^{ev}_{t} \leq P^{ev}_{max}  &&\forall t\\
    \label{equation8d}
    & P^{bess}_{min} \leq P^{bess}_{t} \leq P^{bess}_{max}  &&\forall t
\end{align}
\end{subequations}
being a Quadratic Program, where \( a^{bess}_{t}\) and \( a^{ev}_{t}\) are the internal optimization variables for the BESS and EV that result in the closest feasible, safe actions \( a^{safe, bess}_{t}\) and \( a^{safe, ev}_{t}\) (i.e., the control variables to the assets), starting from the proposed actions \( \tilde{a}^{bess}_{t} \) and  \( \tilde{a}^{ev}_{t} \) originating from the EMSs (i.e., the setpoints from the EMS). Subjecting to the grid exchange (i.e., power balancing) limit and power limits of the BESS and EV charger itself, which can be written out as:

\begin{subequations}
\begin{align}
    \label{equation9a}
    &&&P^{grid}_{t} = - P^{load}_{t} - P^{ev}_{t} + P^{pv}_{t} + P^{bess}_{t}   &&\forall t\\
    \label{equation9b}
    &&&P^{ev}_{t} = f(a^{ev}_{t}, P^{ev}_{max}, SOC^{ev}_{t}) &&\forall t\\
    \label{equation9c}
    &&&P^{bess}_{t} = \gamma^{bess}_{t} \cdot P^{ch}_{t} + (1 - \gamma^{bess}_{t}) \cdot P^{di}_{t} &&\forall t&, \hspace{5pt} \gamma^{bess}_{t} \in \{0,1\} \\
    \label{equation9d}
    &&&P^{ch}_{t} = \begin{cases}
                        a^{bess}_{t} \cdot P^{bess}_{min} &,\text{if} \hspace{5pt} SOC^{bess}_{t} < 1 \\
                        0 &,\text{if} \hspace{5pt} SOC^{bess}_{t} = 1
                    \end{cases} \leq 0 &&\forall t\\
    \label{equation9e}
    &&&P^{di}_{t} = \begin{cases}
                        a^{bess}_{t} \cdot P^{bess}_{max} &,\text{if} \hspace{5pt} SOC^{bess}_{t} > 0 \\
                        0 &,\text{if} \hspace{5pt} SOC^{bess}_{t} = 0
                    \end{cases} > 0 &&\forall t
\end{align}
\end{subequations}
where \(P_t\) are the electrical power variables at time step \(t\) of the grid exchange, non-controllable load, EV charger, solar PV system and BESS, respectively. 
The function, \(f(\cdot)\), in \cref{equation9b} is \cref{ch2_sc3_eq_p_final_soc}. 
A binary decision variable, \(\gamma^{bess}_{t}\), is added to the BESS power equation to avoid simultaneous charging and discharging. 
Hereby, we focus on satisfying the electrical energy balance and the associated grid limit, as no additional constraints are considered in this case study (e.g., ramping rates, minimal run- and downtime).

The \textit{a priori} safe fallback policy \( \pi_{safe} \), which in this case study is activated when the distance to the feasible solution space \(d_{safe}\) is higher than \(1e6\), is simply the self-consumption mode of the BESS together with a re-iteration of the safety layer to determine the maximum EV charging power (as just sending a maximum power setpoint could still violate the grid exchange constraint - and trip the main circuit breaker). 

\subsection{Simulation}
\label{sec:simulation} 

A simulation of the homes is implemented in order to train the TreeC EMS and compare the real experiment with the simulation.
A simulation time step of 15 minutes was used.
The simulation models the behaviour of the BESS, EV charging, PV, electrical load and grid connection and how they react to different control strategies from EMSs.

The simulation model of the BESS is described in \cref{eq:batt_model}.

\begin{equation}
    SOC^{bess}_{t+1} = SOC^{bess}_t + \frac{P^{ch}_{t}*\eta - P^{di}_{t}/\eta^{bess}}{E^{bess}}\Delta t
    \label{eq:batt_model}
\end{equation}
where $SOC_t$ is the SOC at time step $t$, $P^{ch}_{t}$ is the charge power at time step $t$, $P^{di}_{t}$ is the discharge power at time step $t$, $E^{bess}$ is the energy capacity of the BESS, $\eta^{bess}$ is the charge and discharge efficiency of the BESS and $\Delta t$ is the time step duration.
$\eta^{bess}$ is set to 95\% \cite{farhad_introducing_2019} for BESSs 1,2 and 4 and 96\% for BESS 3 as stated in the documentation of the BESS. 
The energy capacity of each BESS is stated in \cref{tab:batt_spec}.

The EV is modelled using  \cref{eq_ev_power_charge,ch2_sc3_eq_p_final_soc,eq_ev_soc_balance}. The PV and electric loads come from the measurements and are included in the simulation. The grid power is calculated with \cref{equation9a}.
To simplify the modelling of the simulations, the limit imposed by the hybrid inverter on the maximum combined power from the PV installation and the BESS is not enforced (see \cref{tab:inverter_spec}).
This simplification is not expected to significantly impact the results of the simulations because the hybrid inverter limits are very rarely reached in practice for the setups used in this study.

The enforced charging behaviour described in \cref{sec:enforced_charging} is closely approximated in the simulation.
The time buffer is calculated using \cref{eq:time_buffer} at each time step.
If following the battery charge or discharge power given by the EMS does not allow reaching the SOC goal at the intermediate time, the battery is charged at the minimum power necessary to meet that goal.

\subsection{Energy management systems}

This section presents the four EMS methods used and the data used to train the MPC, RL and TreeC EMSs.
All EMSs send a setpoint to the BESS and the EV charger every 15 minutes; how they calculate this setpoint is described in this section.

\subsubsection{Training data}
\label{sec:trainingdata}
The MPC, RL, and TreeC EMSs required training data and used three months of data spanning from 2024/01/01 at 00:00 to 2024/04/01 at 00:00.
The dataset included real PV power production, household power consumption described in \cref{sec:consumption} and EV charging sessions from \cref{sec:ev_sessions}.
The MPC utilized this data to train its forecasting models, while the RL agents were pre-trained on the training data with additional power profiles for the BESSs and EV chargers. 
The BESSs and EV charger power profiles were generated as if an RBC EMS was controlling these assets. 
The TreeC EMS used this data in the simulator, described in \cref{sec:simulation}, to construct its decision tree. 

\subsubsection{Rule-based control}

The RBC EMS implemented in this setup consists of default rules commonly found in many devices.
The primary rule for the BESS focuses on self-consumption \enquote{behind-the-meter}. It charges when excess power is available from the PV installation and discharges to meet load demands.
The rule for the EV charger is to always charge the EV immediately at maximum allowed power.
This EMS serves as a benchmark to evaluate the performance of the other EMSs.

\subsubsection{Model predictive control}

MPC is used here as an EMS method that requires a model of the system and forecasts of the system's inputs to optimize control.
The MPC EMS in this setup is implemented as a Mixed-Integer Linear Program, which is solved at each (15-minute) time step.
The horizon of the MPC is variable and ends at the next switching time. The MPC model is solved every 15 minutes.
The variables of the MPC are the power profiles of the BESS and the EV charger, which are the assets it needs to control.

The objective function of the MPC is described in \cref{eq:mpc_obj}.

\begin{equation}
O=\underbrace{\sum_{t\in H}{(Eo_tVo_t-Ei_tVi_t)}}_{\text{day-ahead cost}}+\underbrace{\sum_{t\in H}Eo_tVf}_{\text{offtake extras cost}}+\underbrace{Vp*\max_{{t\in H}} (\frac{Eo_t}{0.25h}, Pp)}_{\text{peak cost}}
\label{eq:mpc_obj}
\end{equation}

Where H is the set of time steps within the horizon of the MPC, and $Pp$ is the previous grid peak power measured for that house when using the MPC EMS. 
At the beginning of the experiment, $Pp$ is set to 2.5 kW.
This objective function penalizes heavily exceeding the previous grid peak power $Pp$.

The MPC system models take as inputs the time of the day, the SOC of the BESS and EV battery, the Belgian day-ahead hourly prices $Vd_t$ until the next switching time, the forecasted power profiles until the next switching time of the PV installation and the electric load and the forecasted final SOC and departure time of the connected EV.

The constraints of the MPC are the same as the ones defined in \cref{eq_ev_power_charge,eq_ev_soc_balance,equation8b,equation8d,equation9a,eq:batt_model}.
However, there is a difference in the model of the EV charging behaviour defined in \cref{ch2_sc3_eq_p_final_soc} as described in \cref{eq:mpc_change_ev}.

\begin{equation}
P^{ev}_t \leq \left\{
\begin{aligned}
&P^{ev}_{max}   &\textrm{if} \quad SOC^{ev}_{\textcolor{red}{t+1}} \leq SOC^{ev}_{cc,cv}  \\
&P^{ev}_{max}  - (P^{ev}_{max}  - P^{ev}_{min}) * \frac{SOC^{ev}_{\textcolor{red}{t+1}} - SOC^{ev}_{cc,cv}}{1 - SOC^{ev}_{cc,cv}}  &\textrm{if} \quad SOC^{ev}_{\textcolor{red}{t+1}} > SOC^{ev}_{cc,cv} 
\end{aligned}
\right.
\label{eq:mpc_change_ev} 
\end{equation}

Calculating the average maximum power of the EV charger over the next time step when the $SOC^{ev}_{cc,cv}$ is exceeded would introduce an exponential relationship to the MPC model.
The linearity of the MPC model is kept by limiting the maximum power of the EV based on the SOC of the EV at the following time step $t+1$ instead.

A constraint requires the modelled SOC of the BESS to be charged to 100\% at the next switching time. A similar constraint is imposed to ensure the EV battery reaches the forecasted final SOC at the forecasted departure time.
This forecasted SOC and departure time can be underestimated, leading to the activation of the enforced charging behaviour (see \cref{sec:enforced_charging}).
For this reason, the approximated enforced EV charging power for the next time step (see \cref{sec:simulation}) is set as the minimum EV power setpoint in the MPC model.
Finally, a constraint requires the modelled grid power not to exceed the maximum grid power of 9.2 kW.

Forecasting of electrical load and PV generation are accomplished using a Light Gradient Boosting Machine (LightGBM) model \cite{ke2017lightgbm}. The model predicts the next 96 quarter-hourly values (totalling 24 hours) using the previous 96 quarter-hourly values, the time of the day and the day of the week. 
An optimal combination of hyperparameters is found using the Tree-structured Parzen Estimator \cite{bergstra2011algorithms}. 
The ranges of hyperparameters used for the optimization are shown in Table \ref{tab:forec_hyperparameters}. 
We refrain from using the weather data provided by external services as an input feature for the PV forecast to avoid possible errors, such as the external server being down for maintenance.
For electrical load prediction, the model is trained using the original data (see \cref{sec:consumption} and therefore the same for the four houses while for PV generation the model is trained on a building-by-building basis. 
The forecasting model for EV charging duration and demands employs k-Nearest Neighbors (k-NN) \cite{cover1967nearest} regression. The k-NN models utilize the initial SOC and the hour of the day as features, leveraging cosine similarity to find the most similar historical patterns. This model has previously shown superiority in data-scarce applications, such as the prediction of EV charging session parameters \cite{GENOV2024121969}. 

\begin{table}[H]
    \caption{Hyperparameter Ranges for LightGBM Model}
    \centering
    \begin{tabular}{@{}l||l@{}}  
        \textbf{Parameter} & \textbf{Range} \\ 
        \hline
        Learning Rate & 0.001, 0.01, 0.1, 0.2, 0.3 \\ 
        Num Leaves & 32, 64, 128, 256 \\ 
        Max Depth & 2, 4, 6, 8, 10 \\ 
        ColSample ByTree & 0.6, 0.8, 1.0 \\ 
        Min Child Samples & 1, 2, 3, 4, 5 \\ 
    \end{tabular}
    \label{tab:forec_hyperparameters}
\end{table}

\subsubsection{Reinforcement learning}
\label{sec:rl}

We formulate the energy managing RL agent as a fully observable discrete-time Markov Decision Process, characterized by the tuple \( \langle S,A,P_a,R_a \rangle \) so that:

\begin{subequations}
\begin{align}
    \label{equation13a}
    S_t = ( P^{load}_t,\ P^{pv}_t,\ SOC^{bess}_t,\ SOC^{ev}_t,\ Vd_t,\ H_t,\ D_t  )& & S_t \in S\\[1em]
    \label{equation13b}
    a^{bess}_t = (a^{bess}_{min} \xrightarrow{} a^{bess}_{max})& & a^{bess}_t \in A  \\
    \label{equation13c}
    a^{ev}_t = (0 \xrightarrow{} a^{ev}_{max})& & a^{ev}_t \in A \\[1em]
    \label{equation13d}
    P_a(s, s') = Pr(s_{t+1} = s' \ | \ s_t = s, a_t = a)& \\
    \label{equation13e}
    R_a(s, s') = - ( \underbrace{Eo_tVo_t-Ei_tVi_t}_{\text{day-ahead cost}} + \underbrace{Eo_tVf}_{\text{offtake extras cost}} ) - z&
\end{align}
\end{subequations}

Where \(P^{load}_t\) is the non-controllable electrical load, \(P^{pv}_t\) the electrical solar in-feed, followed by the SOC of the BESS and EV, \(Vd_t\) the electrical price signal (Belgian day-ahead price), \( H_t \) the hour of the day and \( D_t \) the day of the week, all at time step \( t \) - constituting the state-space \(S\). The continuous action-space \( A \) consists of, the control actions \( a^{bess}_t \) for the BESS and \(a^{ev}_t\) for the EV charger. Furthermore, \( P_a \) signifies a transition probability that exclusively relies on the current state and is unaffected by prior states (in other words, adhering to the Markov Property), for when the system is in a specific state \( s \in S \) at time step \( t \) and takes action \( a \in A \), which would result in the system being in state \( s' \in S \) at the subsequent time step \( t+1 \).

The objective is formulated, that is, the reward function so that when maximizing this function, we minimize the positive version of that function. Hence, we minimize the day-ahead and offtake extras costs (from \cref{eq:tot_cost}, yet calculated over the last 15-minute time step) in EUR and with an additional cost \( z = 1 \) to further shape the reward when the original, uncorrected, predicted action \(\tilde a\) was \textit{expected} to violate constraints. Note that the RL agent maximizes the expected sum of discounted rewards (i.e., cumulative reward).

As the specific RL algorithm, a twin delayed deep deterministic policy gradient agent is used (TD3 \cite{fujimoto_addressing_2018}, that is, twin delayed DDPG) from the \texttt{stable baseline} \cite{raffin_stable-baselines3_2021} implementations with the following hyperparameters:

\begin{table}[H]
    \centering
    \begin{tabular}{l||c}
    \textbf{Parameters} & \textbf{TD3} \\
    \hline
    gamma           &  0.7 \\
    learning\_rate  &  0.000583 \\
    batch\_size     &  16 \\
    buffer\_size    &  1e6 \\
    train\_freq     &  (4, \enquote{step}) \\
    noise\_type     &  normal \\
    noise\_std      &  0.183 \\
    learning\_starts & 96
    \end{tabular}
    \caption{TD3 hyperparameters. The parameters are the result of an optimization study from \citeauthor{ceusters_safe_2023} \cite{ceusters_safe_2023}, yet with an increased \texttt{train\_freq} and reduced \texttt{learning\_starts}.}
    \label{tab:RL_hyperp}
\end{table}

The RL agents (one per house) are pre-trained using Behavioral Cloning (BC) \cite{bain_framework_1999} with the available 3-month historical datasets as described in  \cref{sec:trainingdata}. With a simple random train-test split using 20\% test data and over 10 random seeds, we get the following metrics:

\begin{table}[H]
    \centering
    \begin{tabular}{l||c|c|c|c}
    \textbf{Metric} & \textbf{House 1} & \textbf{House 2} & \textbf{House 3} & \textbf{House 4}\\
    \hline
    R2-score [-]       &0.4746  &0.4997 &0.5837 &0.4617 \\
    MAE [Watt]         & 372    & 344   & 481   & 299\\
    RMSE [Watt]        & 770    & 734   & 855   & 721
    \end{tabular}
    \caption{Pre-trained TD3 BC metrics over 10 random seeds and using 20\% randomly split test data and with: \texttt{epochs} = 3, \texttt{gamma} = 0.7, \texttt{learning\_rate} = 1.0, \texttt{batch\_size} = 64 and \texttt{test\_batch\_size} = 100.}
\end{table}

\subsubsection{TreeC}

TreeC is an EMS method that uses decision trees to control a system.
The decision trees are generated using a metaheuristic algorithm.
The generated decision trees are evaluated using a simulation of the controlled system over a training period.
The metaheuristic algorithm generates better decision trees over the course of the optimization process; the best decision tree is then used as the EMS.
A more detailed description of the TreeC method can be found in \cite{ruddick_treec_2023}.

One decision tree is generated to control the BESS and another one to control the EV charger for each house.
The decision trees used in the experiment are generated using the training data of \cref{equation13a}, the simulator of \cref{sec:simulation} and the objective function described in \cref{eq:tot_cost}.

Two changes have been made to the original TreeC method described in \cite{ruddick_treec_2023}:

\begin{itemize}

\item The metaheuristic algorithm Covariance matrix adaptation evolution strategy seemed to end up in local minima during different training runs regularly.
The covariance matrix adaptation evolution strategy was therefore replaced by the metaheuristic algorithm particle swarm optimization \cite{kennedy_particle_1995}, which obtained better training results.
This study used the particle swarm optimization algorithm implemented in the Python library pygmo \cite{biscani_parallel_2020} with default parameters which corresponds to the canonical particle swarm optimization algorithm described in \cite{poli_particle_2007}.
\item At the end of the training, the best decision tree is pruned by removing all leaves that do not make the objective function score worse than 1\% compared to the unpruned decision tree.
The least used leaves are tested for removal first.
\end{itemize}

The splitting features of the decision trees are the state space of the RL EMS of \cref{sec:rl} and an additional feature representing the day-ahead price shifted to have the minimum price of the ongoing experiment day equal to 0.
This additional feature is calculated by subtracting to the current day-ahead price the minimum price of the experiment day period ranging from the previous switching time to the next switching time.

The actions of the TreeC EMS include charging or discharging the BESS at different power levels and also the possibility to do self-consumption.
The action space is normalized between 0 and 1. 
If the action value is less than 0.1, then it does self-consumption; otherwise, it charges or discharges the BESS at the power level corresponding to the denormalized action value.

The TreeC EMS is trained in simulation with the particle swarm optimization algorithm over 1000 generations and a population of 1000 individuals.
This means 1\,000\,000 decision tree based EMSs are evaluated in simulation on the three months of training data and the best one is then selected.
The population size of 1000 was chosen as it obtained better results for the real-world and non-unimodal benchmarks of the IEEE Competitions in Evolutionary Computation compared to the more historically implemented population sizes of 20-50 individuals \cite{piotrowski_population_2020}.
As in classical TreeC, five different EMSs are generated for each house through training then pruning.
The best-performing EMS of the five is then used in the experiment.

\begin{figure}[H]
    \centering
    \begin{subfigure}[b]{0.45\textwidth}
        \centering
        \includegraphics[width=\textwidth]{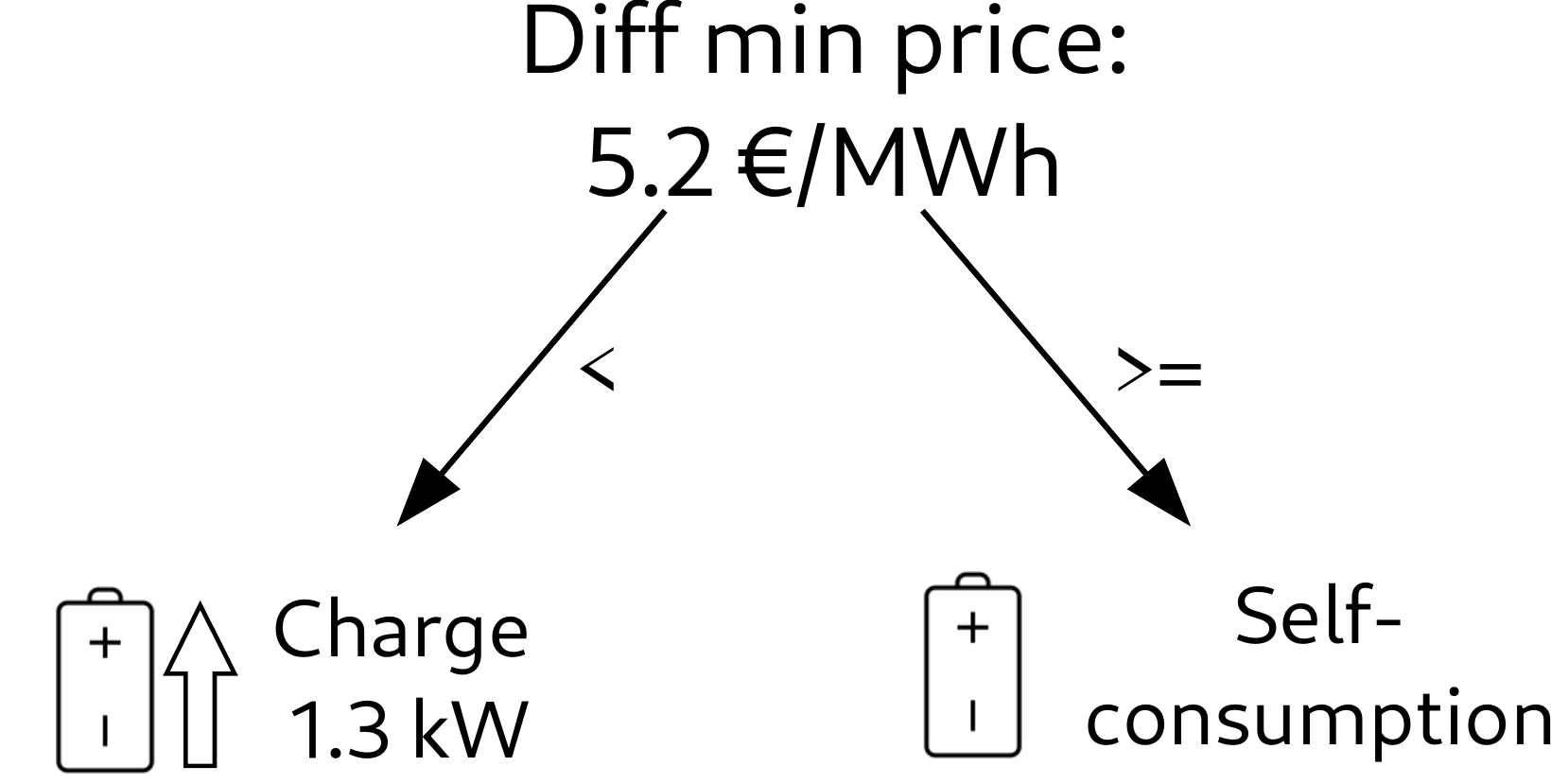}
        \caption{Decision tree for the BESS}
        \label{fig:house1_bess}
    \end{subfigure}
    \hfill
    \begin{subfigure}[b]{0.45\textwidth}
        \centering
        \includegraphics[width=\textwidth]{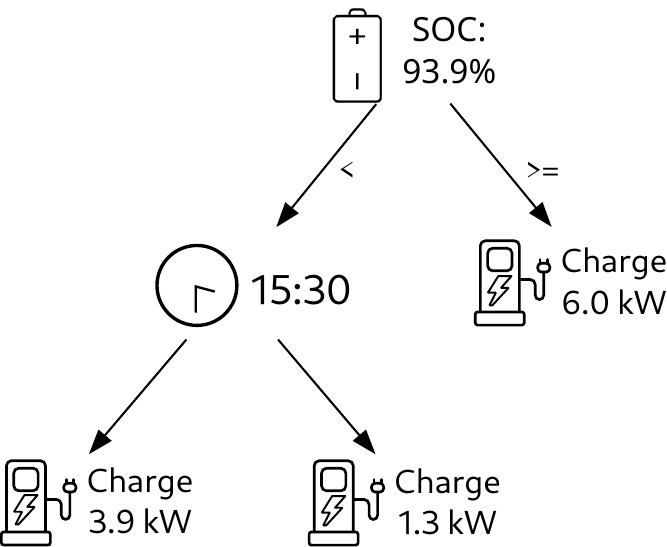}
        \caption{Decision tree for the EV charger}
        \label{fig:house1_ev}
    \end{subfigure}
    \caption{The TreeC EMS obtained after training and pruning for house 1.}
    \label{fig:house1_decision_trees}
\end{figure}

The TreeC EMS obtained after training for house 1 is presented in \cref{fig:house1_decision_trees}. 

The obtained EMS is interpretable.
The BESS charges when the electricity price is close to the minimum price for the experiment day and otherwise does self-consumption. 
The EV charger sets a high charging power when the BESS is almost full; otherwise, it sets a low charging power after 15:30 and an intermediate charging power before 15:30.

\subsection{Experiment execution}

The experiment was conducted from 2024/04/11 at 15:00 to 2024/06/17 at 15:00 to obtain 48 experimental days fit for comparison.
In total, 19 experimental days were not used due to BESSs not reaching 100\% at the switching time, missing data or circuit breaker trips.
The reasons for these failures were the following:

\begin{itemize}
\item Server restart interrupting the execution program (2 days)
\item BESS of house 1 stopped responding until manual intervention (2 days)
\item BESS of house 4 stopped responding when fully charged, a bug fix was applied to limit the SOC to 95\% (1 day)
\item The bug fix for house 4 did not fix the issue. A new fix was applied to perform self-consumption above 95\% SOC (2 days)
\item A circuit breaker from the dynamic loads circuit of house 2 tripped because the breaking capacity was too low for high loads; a circuit breaker with a higher breaking capacity was used instead (6 days)
\item A previously uncaught error caused the experiment program to stop (2 days)
\item The main circuit breaker of house 3 blew because the apparent power of the house was too high for too long (2 days)
\item Missing grid measurements due to a malfunction of the grid meter of house 1 (2 days).
\end{itemize}

Some changes have been made to the experiment during its execution.
As listed above, we encountered an unexpectedly high amount of reactive power from the dynamic loads within our hardware setup (\cref{fig:setup_diagram}), causing a circuit breaker trip. 
This as, the safety layer was formulated using active power rates, essentially assuming a perfect power factor of 1. 
Therefore, halfway through the experiment, we incorporated the reactive power in the constraints (\cref{equation8b} and \cref{equation9a}) - imposing limits on the grid's apparent power instead. 
This did convert \cref{equation9a} into a quadratic constraint, resulting in a Mixed-Integer Quadratically Constrained Quadratic Program. 
However, this now did limit the maximum main circuit breaker current correctly.
Extra measurements were taken on the main grid breaker, and the grid limit of 9.2 kVA apparent power corresponds to approximately 8.7 kW active power. 
The maximum grid power in simulation and the MPC constraints was therefore changed from 9.2 kW to 8.7 kW.
The correction was made on the 2024/5/21 at 15:00.

Two other minor changes were made during the experiment.
The EV SOC input used by the TreeC EMS had a value 100 times too low. This was corrected and had an impact on only one charging session without affecting the performance much. 
The error was also reproduced in the simulation (Corrected 2024/4/15 15:00:00).
The maximum SOC for the BESS of house 4 was limited to 95\% as described in the bullet points above (First correction 2024/04/22 15:00:00, second correction 2024/04/30 15:00:00).

\begin{figure}[H]
    \centering
    \includegraphics[width=\textwidth]{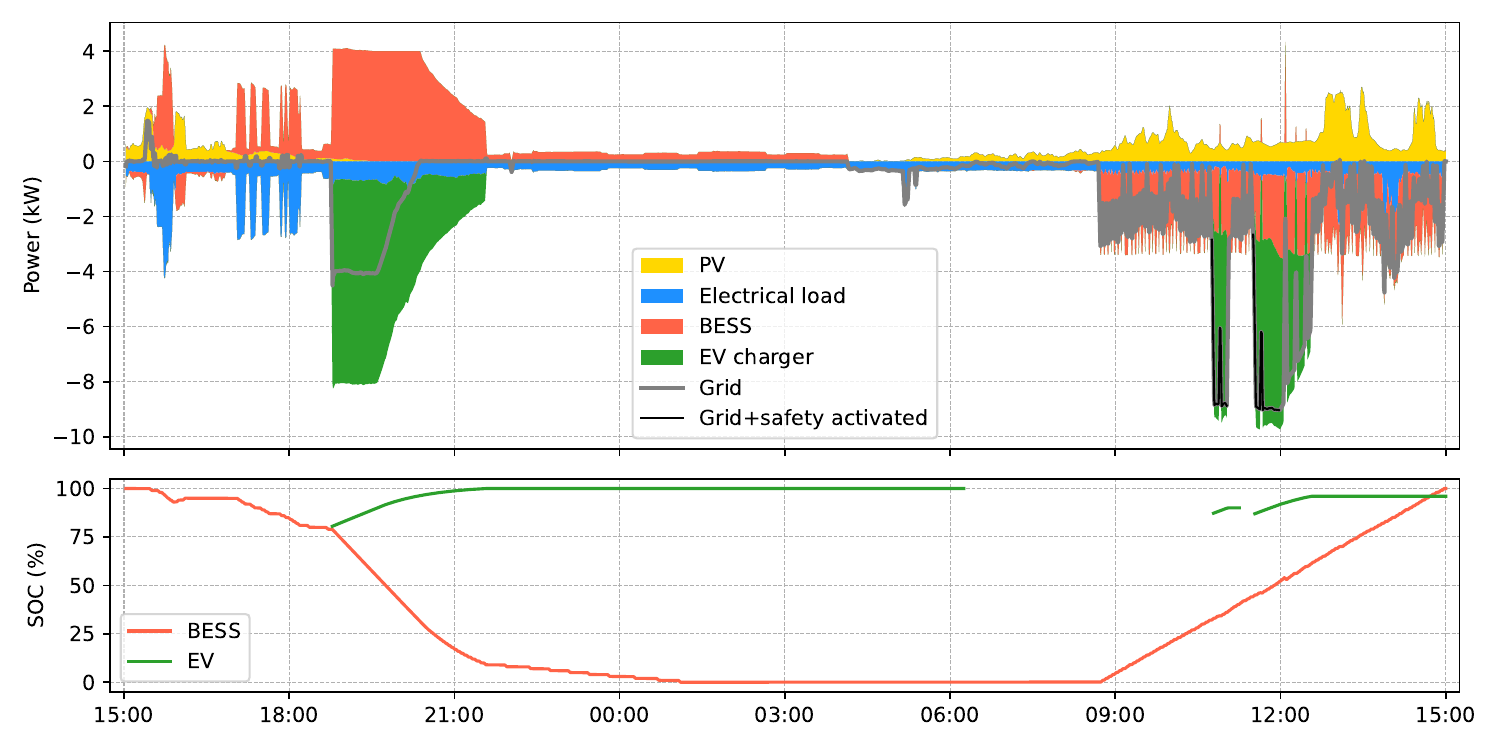}
    \caption{
    Representation of the experiment day starting on the 30th of May at 15:00. The upper plot shows the power profiles of the different assets of house 3 controlled by the RBC EMS. The grid power is represented in grey when the safety layer is not activated and in black when it is. The bottom plot shows the SOC of the BESS and EV. The enforced charging behaviour of the BESS is executed at the end of the experiment day to reach the 100\% SOC goal at 15:00. 
    }
    \label{fig:RBC_example}
\end{figure}

\Cref{fig:RBC_example} shows the RBC EMS controlling house 3 for an experiment day. 
The figure shows the power profiles of the different assets for this experiment day, the SOC of the EV and BESS, the enforced charging behaviour at the end of the day for the BESS and two periods when the safety layer corrected the EV and BESS powers.

\section{Results and discussion}
\label{sec:results}

\subsection{EMS performance comparison}
\begin{figure}[H]
    \centering
    \includegraphics[width=0.8\textwidth]{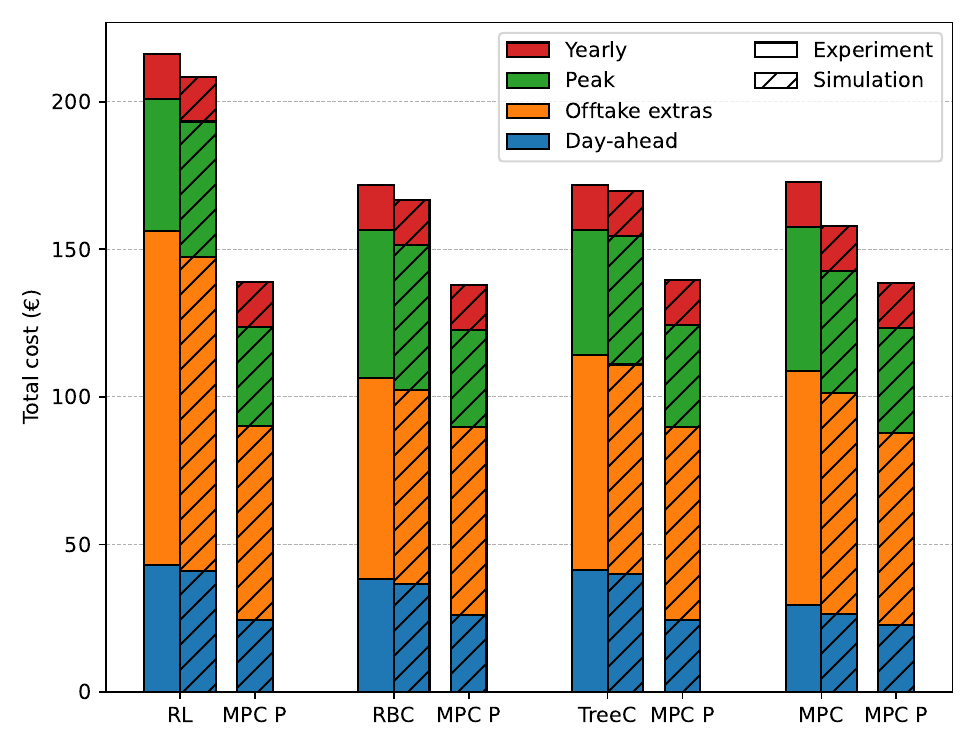}
    \caption{Costs of electricity for each EMS over the 48-day experiment from the real measurements and simulation.
    The total cost of electricity is divided into the four different costs from \cref{eq:tot_cost}.
    In addition to EMSs costs, the simulated costs of an MPC EMS with perfect forecast is also shown with the label MPC P.
    The MPC P costs are simulated using the same set of 48 days and house combinations as the EMS on its left and serve as benchmarks.
    }
    \label{fig:comparison_results}
\end{figure}

The RBC, TreeC and MPC EMSs obtained similar performances in the experiment as shown in \cref{fig:comparison_results}.
The RL EMS obtained a worse performance than the other EMSs, which is expected as the RL performed more exploratory actions than the other EMSs.
For the three other EMSs, them being close is unexpected but is explained in the following sections.

No EMS is close in performance to the MPC P benchmarks, indicating better implementations or methods could still be found.
The MPC P EMSs obtained scores between \euro137.9 and \euro139.5 even though they were evaluated on different day house combinations (i.e. the same day house combinations than the EMS for which they serve as a benchmark). 
The score difference is small, which indicates that the comparison procedure is fair.

\begin{table}[H]
    \label{tab:import_export_res}
    \centering
    \begin{tabular}{l||c|c|c|c}
    \textbf{Result} & \textbf{RL} & \textbf{RBC} & \textbf{TreeC} & \textbf{MPC}\\
    \hline
    Grid imported [kWh]       & 991.7  & 595.3 & 638.6 & 694.9 \\
    Grid exported [kWh]         & 458.3    & 91.3   & 142.2   & 190.5\\
    Net consumption [kWh]        & 533.4    & 504.0   & 496.3   & 504.4
    \end{tabular}
    \caption{Electrical energy metrics for each EMS over the 48-day experiment from the real measurements. Net consumption is defined as the difference between grid-imported and grid-exported energy.}
\end{table}

The results of \cref{tab:import_export_res} show that the RL EMS imported more electricity from the grid compared to the other EMSs, and it also has a higher net consumption. A higher net consumption is most probably caused by a higher use of the battery, which would result in more efficiency energy losses.

\subsubsection{Rule-based control}

The strategy of the RBC is to do as much self-consumption as possible, which means trying not to take electricity off the grid as shown in \cref{tab:import_export_res}.
This strategy is reflected in a comparatively low offtake extras cost, which is proportionally the highest cost category for all EMSs.

\subsubsection{TreeC}

The TreeC EMS obtained the lowest peak cost but couldn't obtain as low offtake extras and day-ahead costs as the other MPC and RBC EMSs.
Looking at the decision trees obtained in \cref{fig:house1_decision_trees,fig:house2_decision_trees,fig:house3_decision_trees,fig:house4_decision_trees}, the low peak cost is explainable by lower charging setpoints of the EVs.

\begin{figure}[H]
    \centering
    \includegraphics[width=0.8\textwidth]{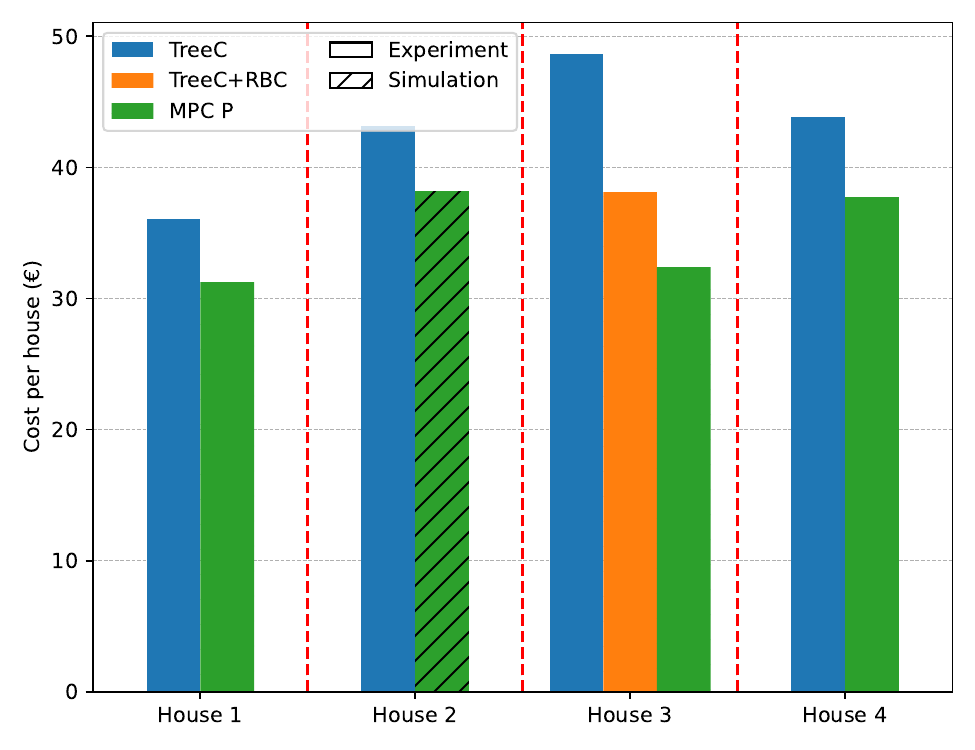}
    \caption{Comparison for each house of the total cost obtained with the TreeC EMS and the benchmark MPC P EMS. 
    For house 3, the TreeC+RBC EMS is added to the comparison and represents the TreeC EMS using the self-consumption control strategy of the RBC EMS to control the BESS.}
    \label{fig:house3_tree_rbc}
\end{figure}

Intuitively, it is surprising that the TreeC EMS did not perform better than the RBC EMS.
\Cref{fig:house3_tree_rbc} shows that the benchmark costs of the MPC P EMS are 10-15\% lower than the TreeC EMS costs for houses 1,2 and 4.
For house 3, though, the MPC P EMS cost is 33\% lower than the TreeC EMS cost, showing a much larger difference than for the other houses.
Looking at the decision trees obtained from the TreeC training (see \cref{fig:house1_decision_trees,fig:house2_decision_trees,fig:house3_decision_trees,fig:house4_decision_trees}), the main difference between houses 1,2 and 4 and house 3 is the BESS strategy.
Houses 1,2 and 4 charge their BESS when the electricity price is very low and otherwise do self-consumption.
House 3, on the other hand, does self-consumption until midnight and then charges the BESS until the switching time at 15:00.
This strategy could well be valid in the training months of January to March when there is less PV production but during the experiment months of April to June, the strategy would lead to more offtake from the grid and, therefore, offtake extras costs compared to the self-consumption strategy of the RBC EMS.
When replacing the TreeC BESS strategy of house 3 with the self-consumption strategy of the RBC EMS in simulation, the MPC P cost is now only 15\% lower for that house as shown in \cref{fig:house3_tree_rbc}.
The TreeC+RBC policy for house 3 would have reduced the total cost of the TreeC EMS by ~\euro10.5, making it better than the RBC EMS over the total cost of the 4 houses (see \cref{fig:comparison_results} for original total costs).

Overall, this shows the importance of using data from a training period that is representative of the experiment or test period.
In this case, the training data was not representative enough of the experiment, which led to the TreeC EMS obtaining a bad control strategy for the BESS of house 3.

\subsubsection{Model predictive control}
\label{sec:mpc_res}

The MPC EMS obtained the lowest day-ahead costs but couldn't obtain as low of a peak and offtake extras costs as the RBC and TreeC EMSs.
The low day-ahead cost is expected as the MPC knows the future Belgian day-ahead prices of electricity and can optimize accordingly.
Intuitively, it is surprising that the MPC did not optimize the other costs efficiently.

\begin{figure}[H]
    \centering
    \includegraphics[width=0.8\textwidth]{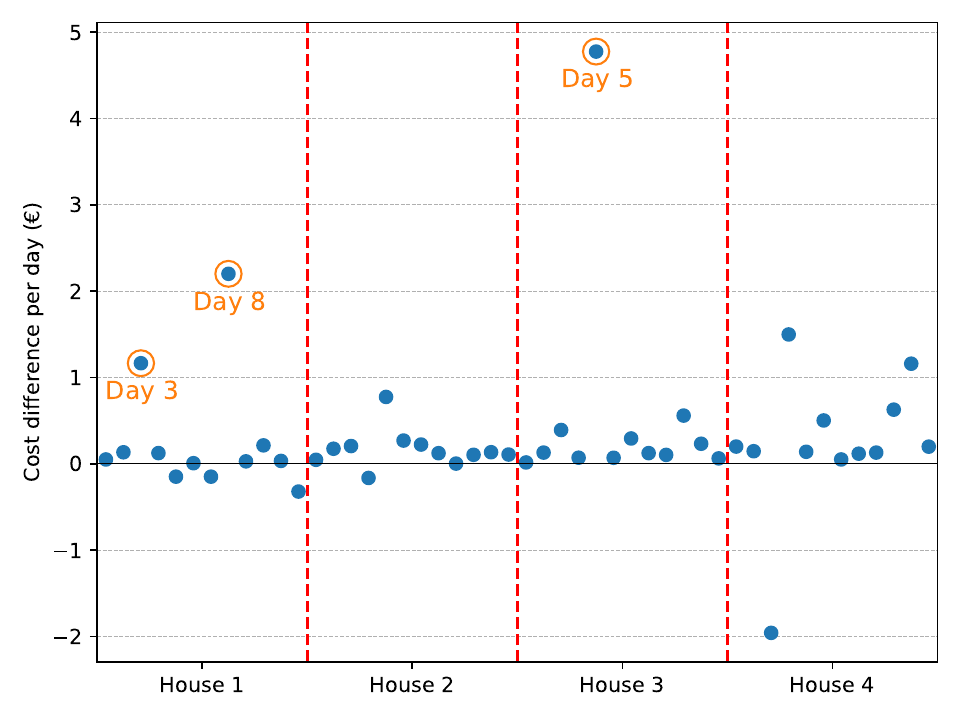}
    \caption{Difference in costs between the experiment and the simulation for the MPC EMS for each day.
    }
    \label{fig:diff_mpc}
\end{figure}

\Cref{fig:comparison_results} shows a bigger cost difference between simulation and reality than the other EMSs and obtained lower peak, offtake extras and day-ahead costs.
Certain days had cost differences exceeding \euro1, as seen in \cref{fig:diff_mpc}.
A deeper analysis was performed to understand these differences on three days that had this cost difference for houses 1 and 3. 
House 4 also had two days when the simulation results were worse than the experiment by more than \euro1.
However, on one day, the simulation outperformed the experiment by approximately \euro2, which mitigates the impact of the two worse days.

In all three instances, the main driver behind the cost differences is the peak cost.

\begin{itemize}
    \item Day 5 of house 3, a difference of \euro4.8: On the first time step of the day, a new short charging session forced the EV charger to charge at maximum power. 
    In the simulation, this peak was avoided by discharging the BESS at the same time, but this did not occur in the experiment.
    The MPC EMS in the experiment decided on the setpoint for the BESS and EV charger without receiving the information that an EV session would start at that time step and, therefore, did not set an appropriate BESS setpoint.
    This is a coding error that would happen when an EV session starts at the first time step of the day.
    \item Day 8 of house 1, a difference of \euro2.2: The peak of the day was higher in the experiment than in the simulation because the MPC set a higher charging power for the EV charger than in the simulation at midnight.
    The MPC EMS forecasted that there would be some PV production at that time because it received the PV production of the previous day in a wrong order.
    This wrong order was due to the device failing to take the grid measurements.
    The device failed to collect data from the \nth{29} of May at 10:45 to the \nth{30} of May at 06:45, and the investigated day started on the \nth{30} of May at 15:00.
    The forecaster of the MPC requested the data from the previous 24 hours, but due to the missing data and the formatting functions not sorting by default on time; the data was received starting from 06:45 until the time of request and then from 24 hours before the time of request until 06:45.
    This mixed order caused the forecaster to predict a PV production between 300W and 400W during nighttime and at a time when the MPC EMS decided to charge the EV at power close to the previous peak, thus obtaining a higher peak in the experiment than in the simulation for the day.
    \item Day 3 of house 1, difference of \euro1.7: Close to the end switch time of the investigated experiment day, the MPC EMS managed in simulation to charge the BESS at 100\% SOC and complete an EV session without increasing the peak cost. In contrast, in the experiment, the peak was increased due to bad management of the EV session and BESS 100\% SOC switching time constraint.
    The error came from a difference in the experiment and simulation MPC. Whenever the forecaster would predict a departure time later than 15:00, the simulation MPC would set the departure time to 15:00 instead. This correction was mistakenly not applied in the experiment MPC.
\end{itemize}

These errors were reproduced in simulation, and the total cost went from \euro157.8 to \euro165. 
This brings the simulation cost closer to the experimental total cost of \euro172.8 by half of the original difference.
A better implementation of the MPC EMS in the experiment would improve its performance.
On the other hand, this shows that the peak cost can penalize the performance of the MPC if there are some errors in the implementation.
EMSs based on less complex models and inputs, such as the RBC and TreeC EMS, did not experience similar errors.
Also, it would have been difficult to notice these errors without a simulation reproducing the MPC behaviour.

\subsubsection{Reinforcement learning}

The RL EMS obtained the worst score, mainly due to a higher offtake extras cost (distribution and taxes) as can be seen in \cref{fig:comparison_results} - which is not unexpected as we observed early-stage learning despite the pre-training, given the new (real) experience tuples.
Moreover, even though the RL agents were pre-trained offline with a 3-month-long historical dataset, each agent-house combination only had an experimental runtime of 1.7 weeks (12 days). When comparing to relevant simulated case studies, more required training time is also observed. For example, \citeauthor{ceusters_model-predictive_2021} \cite{ceusters_model-predictive_2021} reported 95\% MPC-like performance of the RL EMS after \(\sim18.6\) weeks and only matched the MPC EMS after \(\sim2\) years - also using 15-minute timesteps.

\begin{figure}[H]
    \centering
    \includegraphics[width=0.8\textwidth]{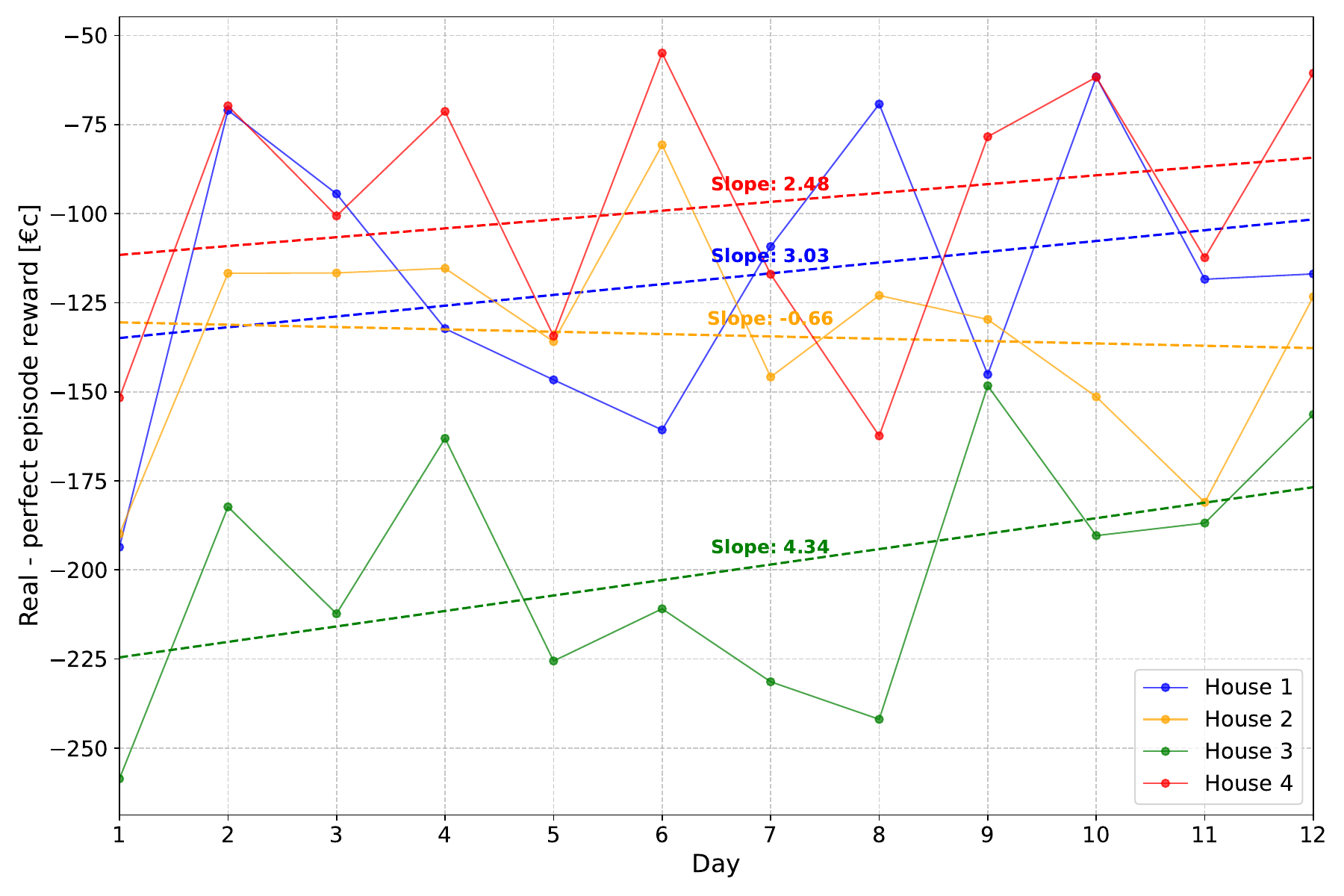}
    \caption{Difference in episodic rewards (in EUR cents) between real RL agents and the simulated MPC with perfect foresight (MPC P) that shows the learning rate and policy improvement of the RL agents. 
    Note that these episodic rewards are without the peak and yearly costs and that even though only 12 episodes are observed per agent, the policy networks are trained every 4 steps (see \cref{tab:RL_hyperp}).}
    \label{fig:RL_rewards}
\end{figure}

The actor and critic loss, together with the episodic reward, presented in \cref{fig:actor_critic} and \cref{fig:RL_rewards} respectively, provide the necessary insights into the learning process and convergence behaviour of the RL agents. The actor loss (\cref{fig:actor_loss}) indicates that the agents benefited from initial pre-training using behaviour cloning, as evidenced by the initially low actor loss values. 
This suggests that the agents started with a good approximation of the policy used in the pre-training data (the RBC policy). 
However, the subsequent increase in actor loss reflects the agents' ongoing policy refinement and exploration during the online learning phase. 
The critic (\cref{fig:critic_loss}) shows a high initial loss, which rapidly decreases as the critic network learns to evaluate the pre-trained policies accurately. 
This quick reduction in critic loss again demonstrates the effectiveness of the behaviour cloning phase (the critic network is fine-tuning its value function estimates rather than learning from scratch). 
Over time, the critic loss stabilizes at lower values, indicating convergence (towards) and fine-tuning of its value estimates.

To further analyze the performance of the RL agents, we compared the real experimentally obtained rewards against a simulated optimal policy (the MPC with perfect foresight, MPC P, over the same day/house combinations as observed with the real RL agents). 
This as, just analyzing the obtained episodic rewards (of any EMS) shows a significantly higher variance simply due to the environment dynamics itself, i.e. the differences between the days - especially if an EV was present or not. 
The difference, shown in \cref{fig:RL_rewards}, therefore provides better insight into how closely the agents approach the optimal policy over time (or simply improve or not, for that matter). 

The analysis reveals varying rates of improvement across different houses. House 3 shows the most significant improvement, with a positive slope of 4.34 (but also has the worst initial policy), indicating substantial learning progress. 
Houses 1 and 4 also demonstrate positive trends with slopes of 3.03 and 2.48, respectively, suggesting effective learning and policy improvement. 
In contrast, House 2 shows a slight negative slope of -0.66, indicating a more stable or potentially stagnating performance. 
We also note that day 11 is particularly bad for house 2, bringing down the trend. 
These results further highlight the impact of the pre-training, which provided the agents with a strong starting point (day 2 compared to day 1), as reflected in their ability to relatively quickly reduce some of the gap to the optimal baseline - given the very low amount (i.e., 1,152) of real steps in the environment.

However, following those trends, the RL agents would still have a considerable amount of total online training time. This temporary loss in performance and thus monetary training cost would need to be set off against the extra (if any) cost of other methods (e.g. modelling time).
For example, in the context of these residential buildings, the absolute monetary difference is relatively low (10.8 EUR on average per house over 12 days compared to the other EMS, which showed similar performance). Assuming a linear improvement of the RL agent from this point forward (i.e. after the 12 days) and assuming the RL agent matches the MPC performance after 2 years (as seen in \cite{ceusters_model-predictive_2021}), this would result in a total training cost of 333.9 EUR per house. Given the additional engineering cost associated with MPC (but also setting up offline training for the tree-based EMS), it seems reasonable to assume that under these conditions, the total cost of ownership of an RL-based EMS would be lower (especially when the RL-based EMS would outperform the other methods after this training period, as seen in \cite{ceusters_model-predictive_2021}) as well - which is not included in this simple calculation). For cases with a higher absolute energy cost (e.g. industrial and commercial sites), this training cost would increase as well - making the RL EMS less appealing. Yet, it is beyond the scope of this paper to provide conclusive evidence regarding the economic viability of the different EMS methods.
Artificially increasing the sampling rate of the RL agents could benefit the total training time and thus cost (i.e. producing more experience tuples over an equal amount of time) - which is proposed here as a direction for future work.
 
\subsection{Safety layer}

\begin{figure}[H]
    \centering
    \includegraphics[width=0.95\textwidth]{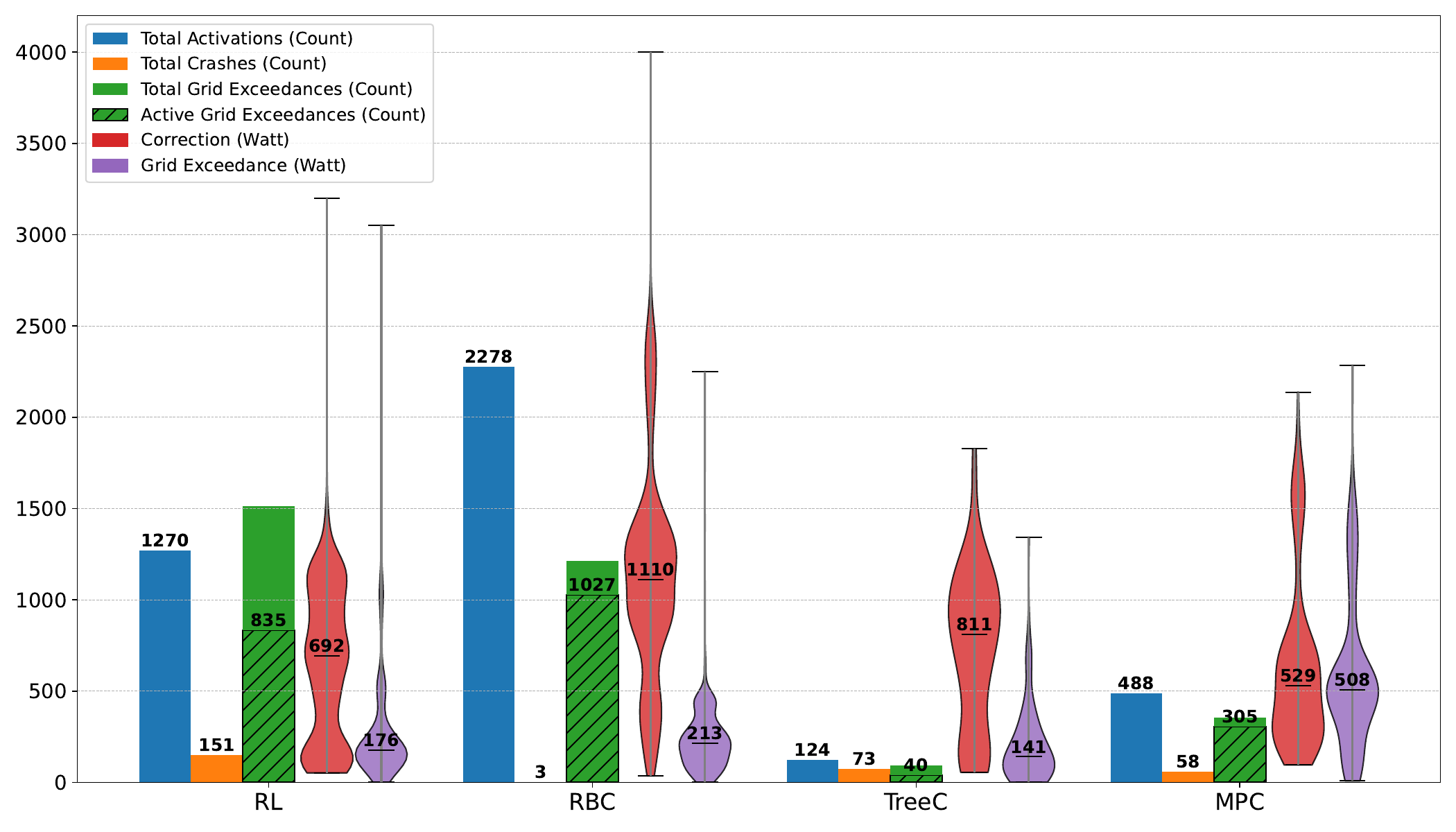}
    \caption{Safety layer results per EMS. Comparing both the number of activations and crashes, as well as how many times the grid limit still was exceeded, together by how much (in Watt) got corrected by the safety layer and how much (in Watt) the grid limit still was exceeded.}
    \label{fig:safetylayer_results}
\end{figure}

The analysis of the safety layer results (\cref{fig:safetylayer_results}) across the different EMS methods shows distinct patterns in the performance and underlying causes of both the safety layer itself and the EMS methods. 

The RBC method experienced the highest number of safety layer activations, primarily due to arguably a flaw in its simple rule-based strategy. 
Given that the BESS was always in self-consumption mode, it would quickly drain itself whenever there would be grid consumption (i.e. trying to avoid any grid exchange) - especially during periods of little to no solar production and at times with high loads. 
The EV charging setpoint then required correcting once the BESS was depleted and a relatively high household load was still present. 
Near the end of the day, the BESS was enforced to charge back to full (see \cref{sec:enforced_charging}). 
If an EV load was still active, this required simultaneously charging the BESS and the EV, which would easily exceed the grid limit (if no or little solar PV infeed is present). 
As a result, the safety layer activates to correct the setpoints preemptively.

The RL method also results in a high number of safety layer activations, as anticipated, due to its exploratory and unconstrained nature. Particularly when attempting to simultaneously charge the BESS and the EV at high power rates (i.e. it not only explored the high power ranges and therefore did not always require the intervention of the safety layer). 
As learning progressed, the RL policies also depleted the BESS quickly at the beginning of the day, resulting in a similar situation as described above near the end of the day, requiring the enforced simultaneous charging of the BESS and EV.

The MPC had an unexpectedly high amount of safety layer activation. Upon reviewing the raw data, it revealed inaccuracies in both the MPC and safety layer constraint formulations - specifically the BESS behaviour at low SOC. 
In reality, the power output of the BESS dropped more quickly than modelled, while both the MPC and the safety layer still \textit{believed} they could discharge the BESS. 
This drop in power for BESS 1 and 2, there was a minimum SOC limit at 5\% and for BESS 1, a discrete power drop of 500 Watt at 10\%. 
Furthermore, BESS 4 only allowed to discharge below 10\% uninterruptedly. 
This means that if the discharging stops at, say, 8\% SOC, the BESS can no longer discharge until it is first charged again above 10\%. 
These specific rules and behaviours were not found in any documentation from the manufacturer. 
In fact, many of the grid exceedances can also be explained by this mismatch in BESS behaviour at low SOC, even the failures of the safety layer to activate (i.e. the unpatterned part of the total grid exceedance count). 
The active grid exceedance count (patterned part) is the number of time steps during which the grid limit was exceeded (active power higher than 9.2 kW) \textit{and} the safety layer was activated, meaning it did not correct enough rather than not at all. 
Nonetheless, the grid exceedance violin plot represents the amount of power (in Watts) over 9.2 kW, regardless of whether the safety layer was activated or not.

The remainder of the caused grid exceedance is due to, in order of importance, load calibration inaccuracies, a time synchronization problem and overall modelling inaccuracies in the safety layer. 
The dynamic load's calibration required a polynomial function that was known to not be constant in time (e.g. initially cold compared to heated up loads and the dynamic loads are installed outside, under the influence of weather variations) and was different per house (while the same function was used for every house), which created an error of approximately 150 - 200 Watts. 
Measurements came from different systems, using different communication protocols and having different (internal) sampling rates, leading to time differences between measurements resulting in sometimes large errors in magnitude but short in time (e.g. when solar PV infeed suddenly drops because of cloud coverage or when ramping up or down the BESS, EV and household load). 
And lastly, the constraint functions themselves are not perfectly accurate, even besides any inaccuracy at upper and lower SOC ranges of the BESS and EV - however, the observed error in normal operational ranges is far lower here (approximately 10 - 20 Watts, with BESS 3 having a considerable settling time of 35 - 40 seconds temporarily resulting in higher observable errors). 
The total summation of grid exceedance in Watt-hour is 593.9 Wh for the RL EMS, 408.4 Wh for the RBC, 27.1 Wh for TreeC and 295.5 Wh for the MPC.

While the TreeC EMS by far has the lowest amount of activations and grid limit exceedances, it is also not perfect as at a rare and short period (approximately 30 min) of high non-controllable household consumption (5 kW), it still required correcting by the safety layer.

Finally, no distinct cause could be determined for any of the safety layer solver crashes (e.g. no measurement gaps and no clear infeasibilities), which leads us to believe that they are mainly due to solver instability. Note that the safety layer ran slightly less than 829,440 times per EMS (every 5 seconds for 12 days on 4 houses, besides a couple of initialization time steps) over the course of the experiment. Hence, the amount of crashes is considered very low (99.97\% uptime). We want to emphasize that these are (limited) crashes of the safety layer solver, not of the EMS solvers.

These results underline both the error-prone nature (possible human modelling errors) and the requirement of accurate constraint functions used in the safety layer. Nonetheless, given that only a subset of equations is used in the safety layer (only those relevant for the constraints), this limitation is lower than, e.g. compared to MPC - that requires more equations \textit{a priori} and so the chance of errors and inaccuracies is higher. For example, \citeauthor{ceusters_adaptive_2023} \cite{ceusters_adaptive_2023} mitigated this limitation by making the constraint functions in the safety layer itself adaptive.

\subsection{Difference experiment vs simulations}

The following section analyses differences between simulation and experiment in addition to the ones already pointed out specifically for the MPC EMS in \cref{sec:mpc_res}.

The experiment runs in real-time while the simulation approximates this behaviour on a 15-minute time step.
In the experiment, the grid sensor measures both imported and exported energy for the same time step, while in simulation, there is only one net consumption value per time step.
The net consumption is the difference between the imported and exported energy from the grid.
Calculating the cost for one time step using both the imported and exported energy instead of the net consumption is always more expensive because offtake extras costs get removed when doing the energy difference for net consumption.
\Cref{tab:energy_diff} shows the score obtained in the experiment for each EMS when using the import and export energy values as well as the score obtained when using the net consumption to understand if the approximation of the simulation has a big impact on the results.

\begin{table}[H]
    \centering
    \begin{tabular}{l||c|c|c}
    \textbf{EMS} & \begin{tabular}{@{}c@{}}\textbf{Experiment} \\ \textbf{(import/export)}\end{tabular} & \textbf{ } \begin{tabular}{@{}c@{}}\textbf{Experiment} \\ \textbf{(net consumption)}\end{tabular} & \begin{tabular}{@{}c@{}}\textbf{Simulation} \\ \textbf{(net consumption)}\end{tabular}\\
    \hline
    RL & \euro216.1 & \euro207.2 &  \euro208.6 \\
    RBC & \euro171.7  & \euro171.2 & \euro166.6 \\
    TreeC & \euro171.7 & \euro170.7 & \euro169.9 \\
    MPC & \euro172.8 & \euro170.1 & \euro157.8 \\
    \end{tabular}
    \caption{Costs of electricity obtained for each EMS in the experiment when taking the export and import energy values and when calculating net consumption.}
    \label{tab:energy_diff}
\end{table}

The cost difference between import/export and net consumption is larger for the RL and MPC EMSs than for the TreeC and RBC EMSs.
A major behavioural difference between the MPC and RL EMSs with the TreeC and RBC EMSs is that the former control the BESS' setpoint while the latter often or exclusively do self-consumption.
Doing self-consumption allows the BESS to adapt to real-time changes and has no or very little imported and exported energy when the BESS is not full or empty.
Giving a fixed setpoint to the BESS for a whole time step, on the other hand, does not allow the BESS to adapt to real-time changes and can lead to both imported and exported energy during the same time step.

A solution to this problem could be to require the BESS to maintain the grid power at a certain setpoint instead of keeping a fixed BESS power for the whole time step.
Further simulation and experimentation would be needed to confirm if this is a good solution.

Both the MPC differences from \cref{sec:mpc_res} and the grid energy exchange calculation difference from this section explain some differences between the experiment and the simulation.
Other differences observed between experiment and simulation but more difficult to quantify in terms of cost include:
\begin{itemize}
    \item BESS capacities and efficiencies not being exactly the same as on the specification sheet.
    \item BESSs 1 and 2 having a clear minimum soc of 5\%.
    \item BESS 4 having a variable minimum SOC, an upped soc of 95\% to avoid the BESS not responding and a recurrent jump from ~85\% to 95\% SOC due to internal recalibration.
    \item The real-time safety layer of the experiment is approximated on a 15-minute time step in the simulation.
    \item The hybrid inverter's power limit is not modelled in the simulation.
\end{itemize}
With the MPC errors of \cref{sec:mpc_res} reproduced in simulation, the difference in total cost between experiment and simulation ranges from 1\% to 4.5\% for all EMSs. When calculating the experiment's total cost based on the net consumption, this range goes down to between  0.5\% to 3\% difference.
The remaining difference between the experiment and simulation can be explained by the points listed above and some other unnoticed differences but, in total, they do not account for a large difference in costs. 
As stated in \cref{sec:hardware}, the electrical loads and possible power losses of the system are not separately measured; therefore, the simulation uses data combining the two.
The possible power losses could depend on the BESS and EV usage, which would not be reflected in the simulation and needs to be further investigated in future work.
Nonetheless, the simple BESS model (see \cref{eq:batt_model}) and the capacity and efficiency parameters provided by the BESS manufacturers (see \cref{tab:batt_spec}) were sufficiently accurate to perform meaningful simulations.

\section{Conclusion}
\label{sec:conclusion}
This paper presented the results of the real-world experimental validation of 2 novel machine learning EMSs: 1) being RL in combination with a safety layer that can handle any constraint type and that can utilize a safe fallback policy when available (i.e., \texttt{OptLayerPolicy} \cite{ceusters_adaptive_2023}) and 2) being a method using a metaheuristic algorithm to generate an interpretable control policy modelled as a decision tree (i.e., TreeC \cite{ruddick_treec_2023}). 
This, together with the comparison between an MPC, a Safe RL, an RBC and an explainable Tree-based EMS on any case study, simulated or otherwise, using a novel evaluation procedure, aimed to be as fair as possible. We come to the following conclusions:

\begin{itemize}
    \item The RBC, TreeC and MPC EMSs obtained a similar economic performance with only a 0.6\% difference in cost between the three. The results highlight the importance of representative training data for the TreeC EMS and expert implementation for the MPC EMS to obtain better performances in real implementations;
    \item The \texttt{OptLayerPolicy} safety layer allows to safely train an RL agent online in the real world, given an accurate constraint function formulation - which remains error-prone, also with MPC;
    \item A pre-trained, yet model-free, RL agent still has a considerably long training time (even with hyperparameters chosen for a higher learning rate) and, therefore, monetary training cost that would need to be set off against the extra (if any) cost of other methods (e.g. modelling time);
    \item The TreeC method showed the safest operational performance (27.1 Wh total grid exceedance compared to 593.9 Wh for RL) in the presented experimental case study, yet does require a simulation model \textit{a priori} and extensive offline training. It is also less error-prone yet requires detailed domain knowledge to predetermine its utility and its operational safety (given it remains unconstrained during deployment without a safety layer);
    \item A real-time safety layer can be beneficial for multiple (optimal) control methods and could streamline the overall (hierarchical) control architecture (given the lack of PID controllers in the experimental setup);
    \item With the MPC experiment errors reproduced in simulation, the difference in costs between experiment and simulation ranges from 1\% to 4.5\% for all EMSs.
    This difference is in part explained by the simulation using a single net consumption value per time step for the energy exchanged with the grid, while the experiment measures both imported and exported energy per time step. 
    Overall, the simulations rather accurately represented reality in terms of cost and were essential to finding the MPC errors and TreeC training data issues;
\end{itemize}

Finally, we propose the following directions for future work:

\begin{itemize}
    \item Investigate if artificially increasing the sampling rate of the RL agents would benefit the overall learning rate to produce more experience tuples more quickly in time (i.e. more time steps over an equal amount of time);
    \item Robustness comparison against faulty and noisy measurements or observations and the requirements (if any) for state estimations;
    \item Assessment of different hierarchical control architectures on optimality, computational complexity, modelling requirements, adaptivity and constraint (incl. Lyapunov functions) handling;
    \item Investigate whether EMSs that rely on more complex models and inputs more frequently experience a performance decline when transitioning from simulation to real-world implementation, as observed with the MPC EMS in this study. 
    \item Re-evaluate the different EMSs with suggested improvements, namely more representative training data for TreeC EMS, fix the errors of the MPC EMS and use the battery to follow a grid power instead of fixing a battery power for one time step.
\end{itemize}


\section*{CRediT authorship contribution statement}

\sloppy \textbf{Julian Ruddick:} Conceptualization, Methodology, Software, Validation,  Formal analysis, Investigation,  Data Curation, Writing - Original Draft, Visualization. \textbf{Glenn Ceusters:} Conceptualization, Methodology, Software, Validation,  Formal analysis, Investigation,  Data Curation, Writing - Original Draft, Visualization, Funding acquisition. \textbf{Gilles Van Kriekinge:} Conceptualization, Methodology, Writing - Review \& Editing. \textbf{Evgenii Genov:} Conceptualization, Methodology, Writing - Review \& Editing. \textbf{Cedric De Cauwer}  Writing - Review \& Editing. \textbf{Maarten Messagie:} Supervision, Funding acquisition. \textbf{Thierry Coosemans:} Supervision, Funding acquisition.

Julian Ruddick and Glenn Ceusters contributed equally to this work and share the first authorship. The order of these authors was determined by a Space Invaders duel, with the highest score winning.

\section*{Declaration of competing interest}
The authors declare that they have no known competing financial interests or personal relationships that could have appeared to influence the work reported in this paper.

\section*{Data availability}

The data, results and code to reproduce the simulation results are available on the following public GitHub repository: \url{https://github.com/EVERGi/real_validation_saferl_treec_paper}.

\section*{Acknowledgement}

This work has been supported by the ECOFLEX project funded by FOD Economie, K.M.O., Middenstand en Energie, by the ICON project OPTIMESH (FLUX50 ICON Project Collaboration Agreement - HBC.2021.0395) funded by VLAIO and by the Baekeland project SLIMness (HBC.2019.2613) funded by ABB n.v. and VLAIO in equal parts.

\appendix
\newpage
\section{Hardware specification of the setup}
\label{sec:appendix_hardware}

\begin{table}[H]
    \centering
    \footnotesize\begin{tabular}{c|l|l|c|c}
        \shortstack{\textbf{House}} & \shortstack{\textbf{BESS model}} & \shortstack{\textbf{Capacity}} & \shortstack{\textbf{Max} \\ \textbf{charge}} & \shortstack{\textbf{Max} \\ \textbf{discharge}} \\
        \hline
        1 & BYD HVS 5.1 & 5.12 kWh & 3.2 kW & 3.2 kW \\ 
        2 & Huawei LUNA2000-5-EO & 5 kWh & 2.5 kW & 2.5 kW \\ 
        3 & D-Centralized LF2 & 15.3 kWh & 3.0 kW & 4.0 kW \\ 
        4 & Pylontech US2000B & 3.55 kWh & 1.7 kW & 2.5 kW \\ 
    \end{tabular}
    \caption{BESS specifications}
    \label{tab:batt_spec}
\end{table}

\begin{table}[H]
    \centering
    \footnotesize\begin{tabular}{c|l|l|c}
        \shortstack{\textbf{House}} & \shortstack{\textbf{Inverter model}} & \shortstack{\textbf{Assets}} & \shortstack{\textbf{Max} \\ \textbf{AC power}} \\
        \hline
        1 & Fronius Primo GEN24 3.0 & PV + BESS & 3.0 kW \\ 
        2 & Huawei SUN2000-5KTL-M1 & PV + BESS & 5.0 kW \\ 
        3 & Sermatec SMT-5K-TL-LV & PV + BESS & 6.0 kW \\ 
        4 & SMA Sunny Boy SB2.5-1VL-40 & PV & 2.5 kW \\ 
        4 & Victron Multiplus-II GX 48/3000/35-32 & BESS & 2.5 kW \\ 
    \end{tabular}
    \caption{Inverter specifications}
    \label{tab:inverter_spec}
\end{table}


\begin{table}[H]
    \centering
    \footnotesize\begin{tabular}{l|c|c|c}
        \shortstack{\textbf{House}} & \shortstack{\textbf{PV peak power}} & \shortstack{\textbf{AC circuit}} & \shortstack{\textbf{House breaker}}  \\
        \hline
        1 & 3.4 kWp & mono phase & 9.2 kVA\\ 
        2 & 5.6 kWp & three phase & 17.2 kVA\\ 
        3 & 3 kWp & mono phase & 9.2 kVA \\ 
        4 & 2.6 kWp & mono phase & 9.2 kVA \\ 
    \end{tabular}
    \caption{Additional specifications}
    \label{tab:pv_spec}
\end{table}

\section{Actor and critic plots of the RL agents}
\label{sec:appendix:actor_critic}

\begin{figure}[H]
    \centering
    \begin{subfigure}[b]{0.8\textwidth}
        \centering
        \includegraphics[width=\textwidth]{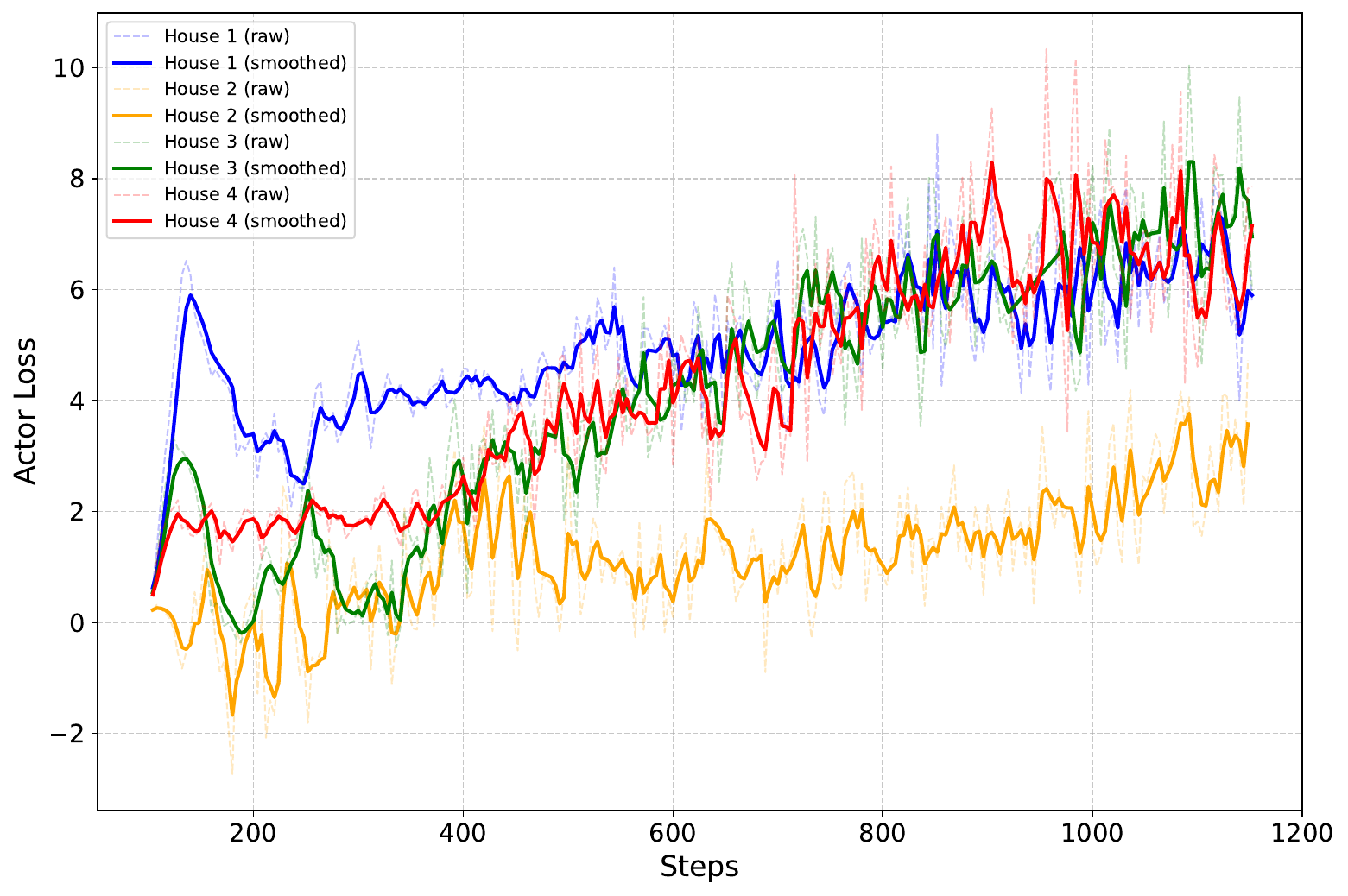}
        \caption{The actor loss over training steps for each house. The initial low actor loss indicates effective pre-training through behaviour cloning, while the increasing trend suggests ongoing policy refinement and exploration during online learning.}
        \label{fig:actor_loss}
    \end{subfigure}
    \hfill
    \begin{subfigure}[b]{0.8\textwidth}
        \centering
        \includegraphics[width=\textwidth]{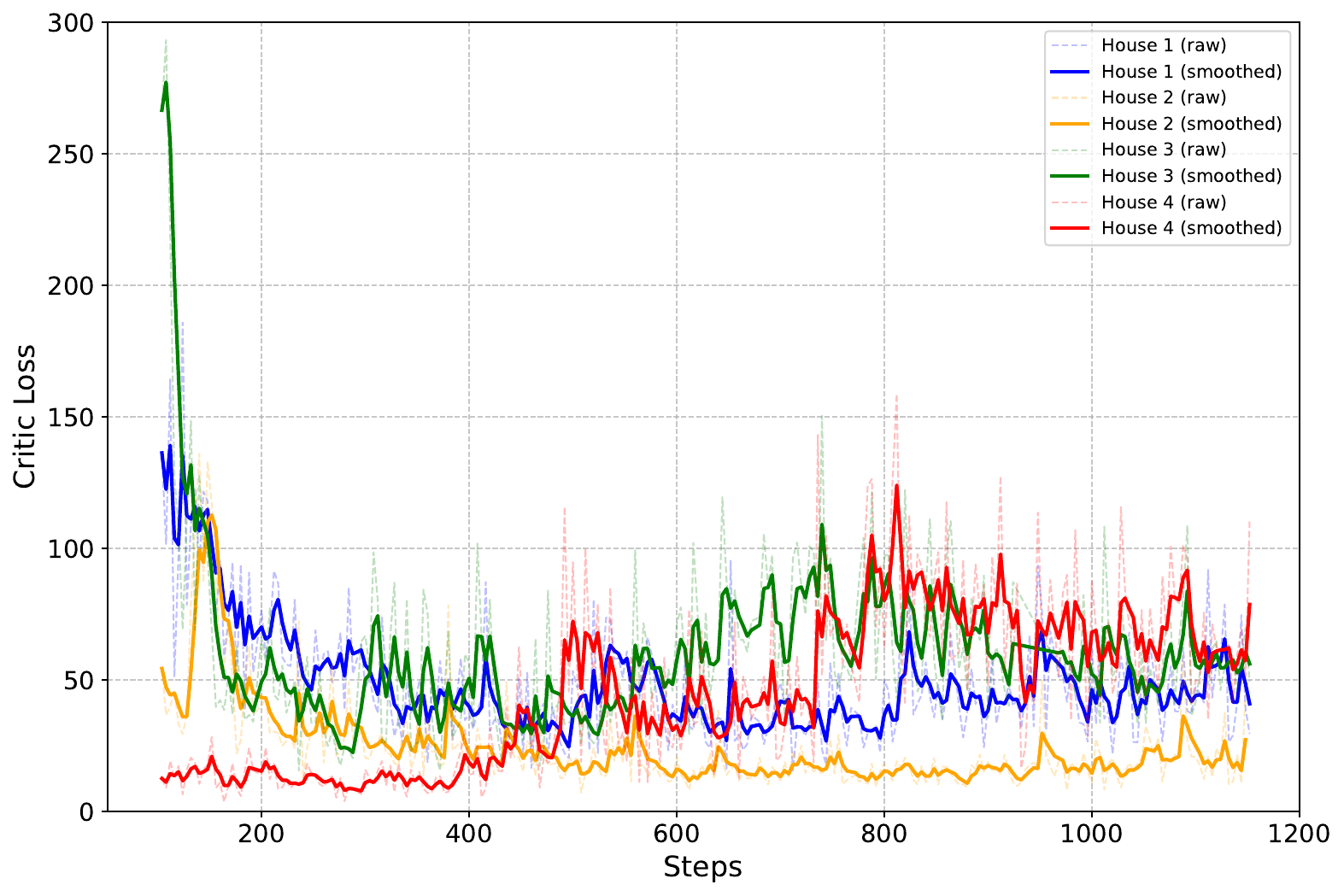}
        \caption{The critic loss over training steps for each house. The high initial critic loss and its rapid decrease demonstrate the critic learning to evaluate the pre-trained policies accurately. The stabilization at lower values over time indicates convergence and fine-tuning of the policies.}
        \label{fig:critic_loss}
    \end{subfigure}
    \caption{RL agents actor and critic loss per house.}
    \label{fig:actor_critic}
\end{figure}

\section{TreeC EMSs for houses 2,3 and 4}

\begin{figure}[H]
    \centering
    \begin{subfigure}[b]{0.45\textwidth}
        \centering
        \includegraphics[width=\textwidth]{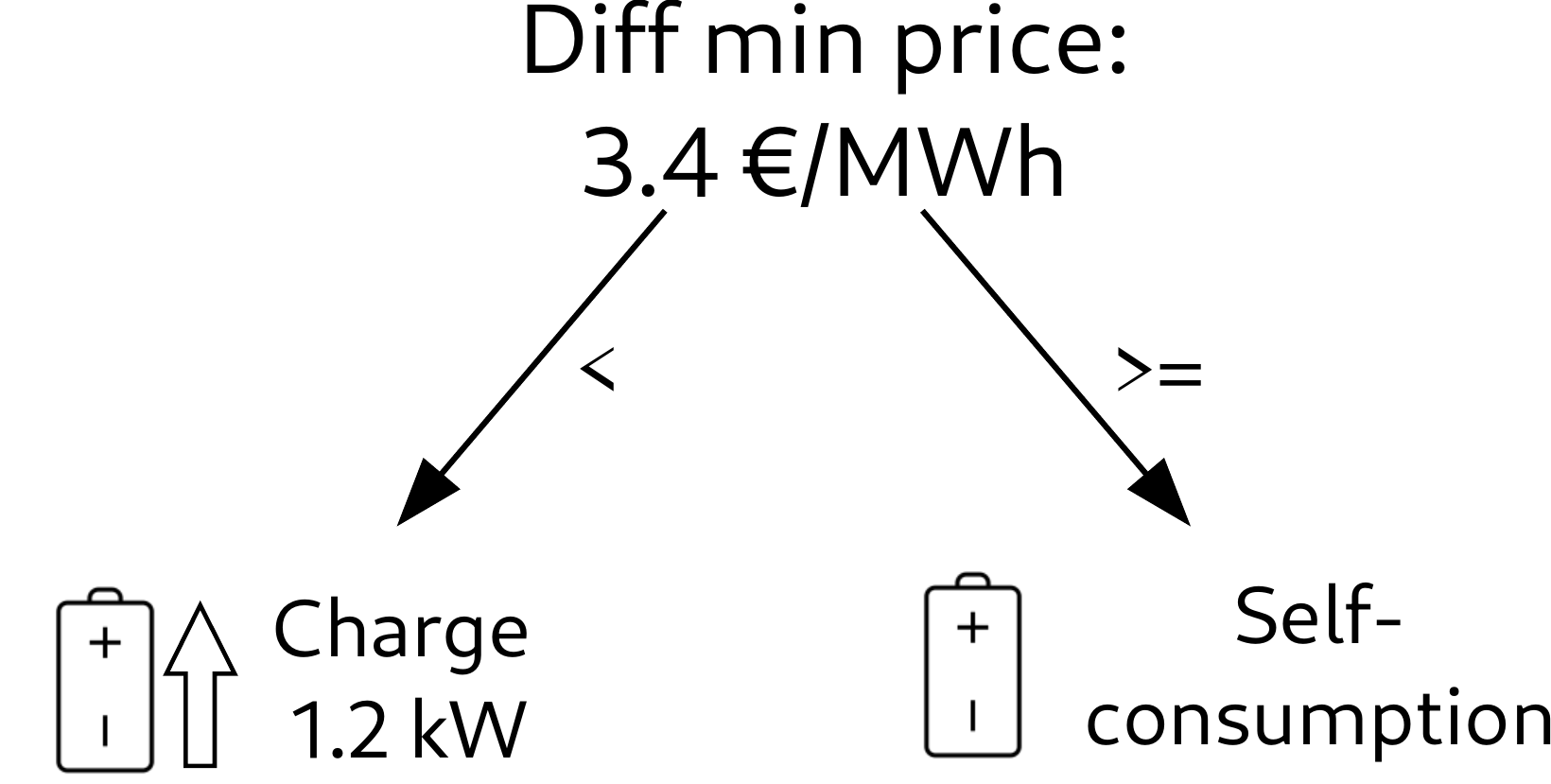}
        \caption{Decision tree for the BESS}
        \label{fig:house2_bess}
    \end{subfigure}
    \hfill
    \begin{subfigure}[b]{0.45\textwidth}
        \centering
        \includegraphics[width=\textwidth]{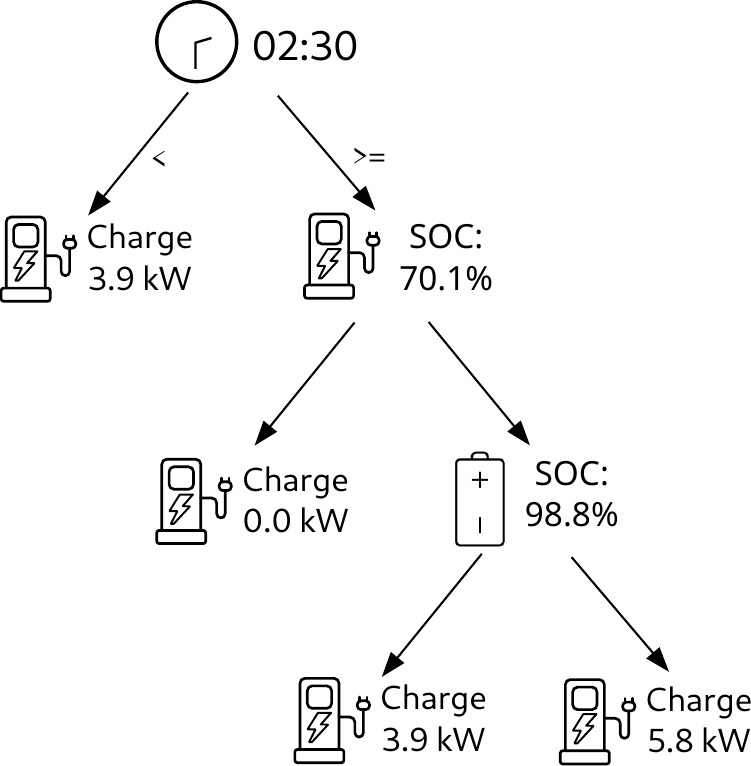}
        \caption{Decision tree for the EV charger}
        \label{fig:house2_ev}
    \end{subfigure}
    \caption{The TreeC EMS of house 2.}
    \label{fig:house2_decision_trees}
\end{figure}
\begin{figure}[H]
    \centering
    \begin{subfigure}[b]{0.45\textwidth}
        \centering
        \includegraphics[width=\textwidth]{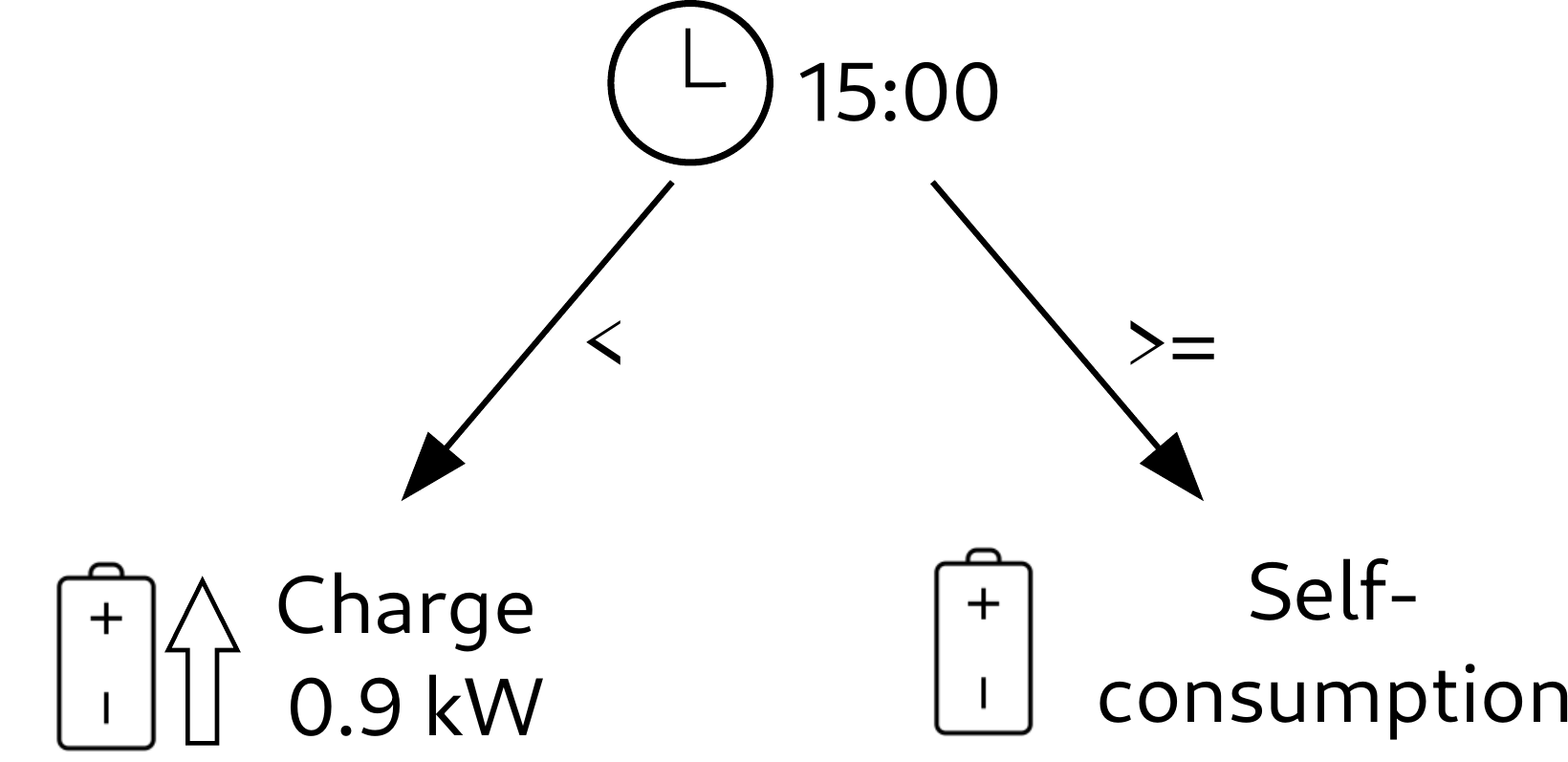}
        \caption{Decision tree for the BESS}
        \label{fig:house3_bess}
    \end{subfigure}
    \hfill
    \begin{subfigure}[b]{0.45\textwidth}
        \centering
        \includegraphics[width=\textwidth]{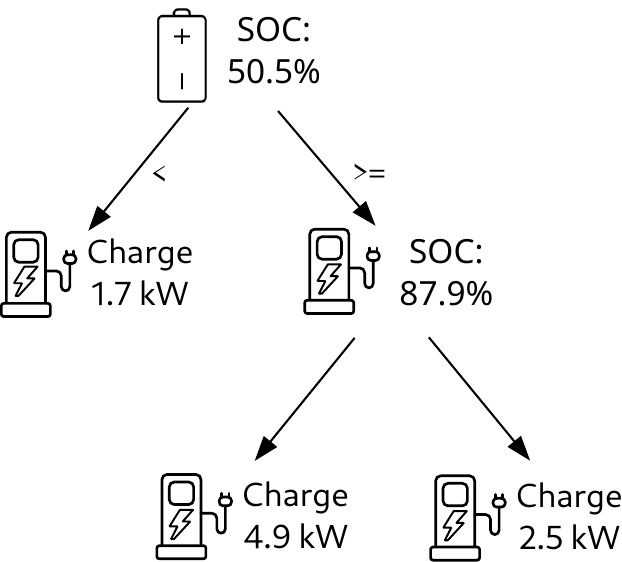}
        \caption{Decision tree for the EV charger}
        \label{fig:house3_ev}
    \end{subfigure}
    \caption{The TreeC EMS of house 3.}
    \label{fig:house3_decision_trees}
\end{figure}

\begin{figure}[H]
    \centering
    \begin{subfigure}[b]{0.45\textwidth}
        \centering
        \includegraphics[width=\textwidth]{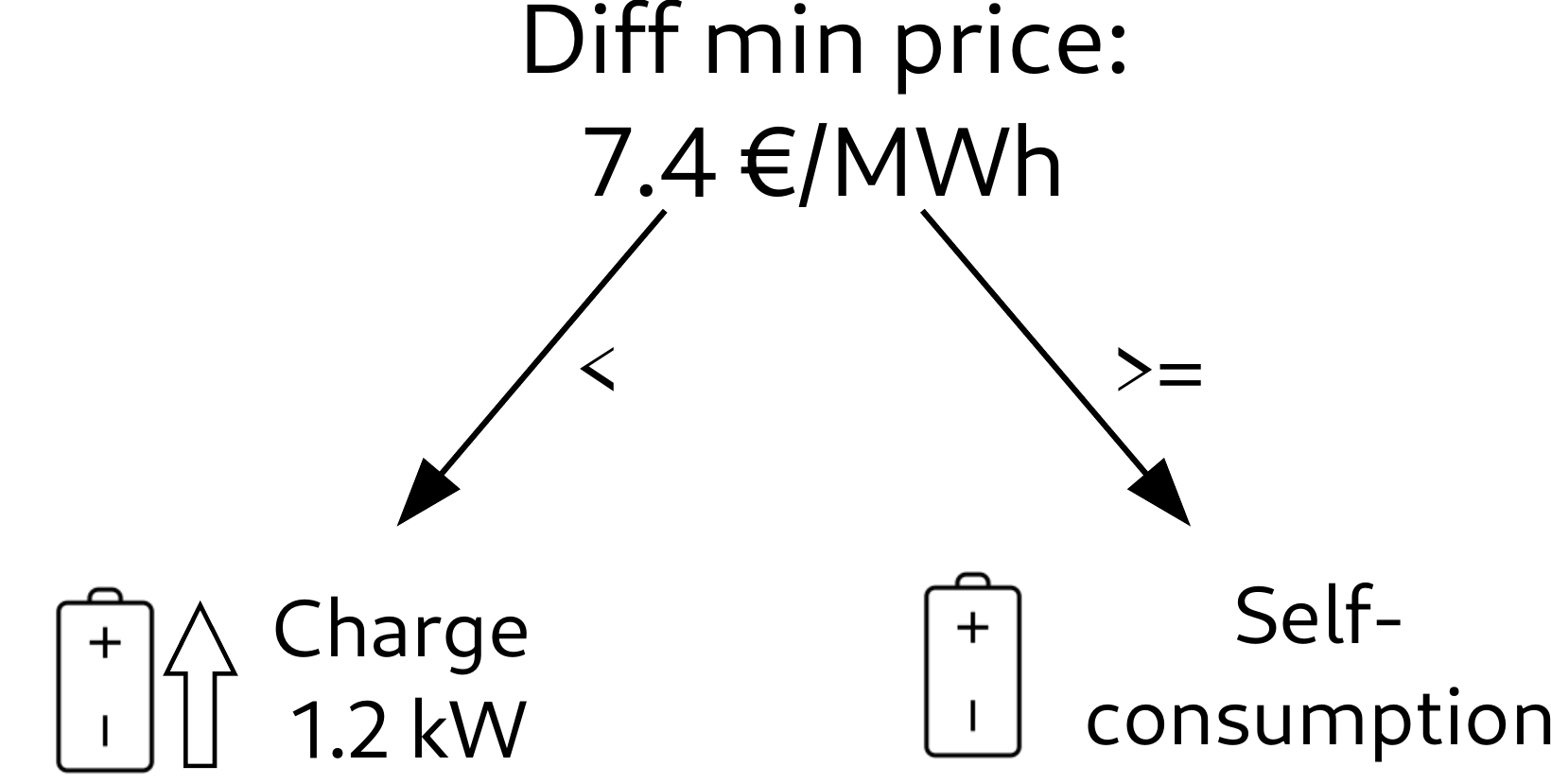}
        \caption{Decision tree for the BESS}
        \label{fig:house4_bess}
    \end{subfigure}
    \hfill
    \begin{subfigure}[b]{0.45\textwidth}
        \centering
        \includegraphics[width=\textwidth]{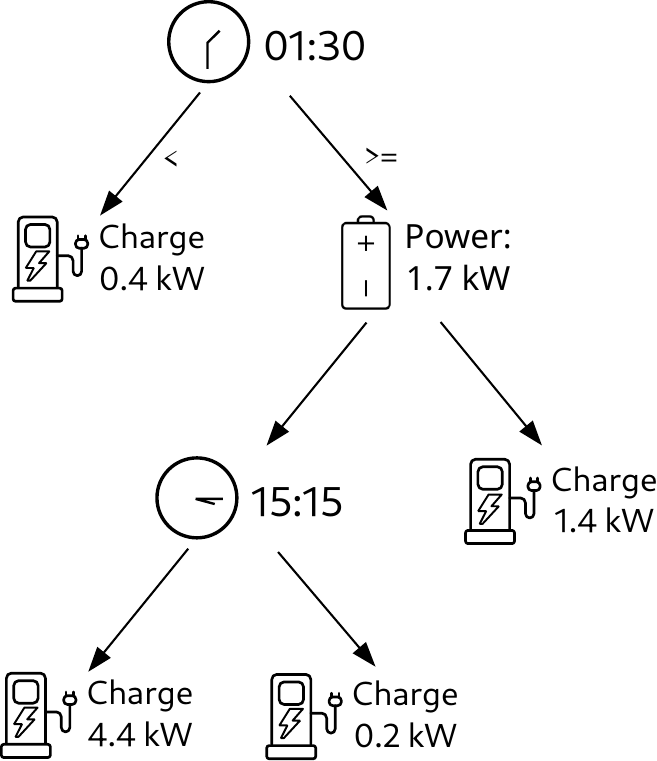}
        \caption{Decision tree for the EV charger}
        \label{fig:house4_ev}
    \end{subfigure}
    \caption{The TreeC EMS of house 4.}
    \label{fig:house4_decision_trees}
\end{figure}


\bibliographystyle{unsrtnat} 
\bibliography{shared_references_jul, references_evgeny}





\end{document}